\documentclass[fleqn,usenatbib]{mnras}

\usepackage{newtxtext,newtxmath}

\usepackage[T1]{fontenc}
\usepackage{ae,aecompl}

\usepackage{graphicx}	
\usepackage{amsmath}	
\usepackage{amssymb}	

\usepackage{rotating, graphicx}
\usepackage{subfig}
\usepackage{pdflscape}
\usepackage{natbib}
\usepackage{amsmath}
\usepackage{courier}
\usepackage{color, colortbl}
\usepackage{float}
\restylefloat{table}
\usepackage{footnote}
\usepackage[T1]{fontenc}
\usepackage{aecompl}
\usepackage{amssymb}
\usepackage{rotating}
\usepackage{mathtools}
\usepackage{enumitem,booktabs,cfr-lm}
\usepackage[referable]{threeparttablex}
\renewlist{tablenotes}{enumerate}{1}
\usepackage{multirow}

\definecolor{Gray}{gray}{0.9}

\makeatletter
\setlist[tablenotes]{label=\tnote{\alph*},ref=\alph*,itemsep=\z@,topsep=\z@skip,partopsep=\z@skip,parsep=\z@,itemindent=\z@,labelindent=\tabcolsep,labelsep=.2em,leftmargin=*,align=left,before={\footnotesize}}
\makeatother

\def\aap{Astronomy \& Astrophysics}

\def\apjs{The Astrophysical Journal Supplement}
\def\apjl{ApJL}

\title[Feedback at low metallicities]{Feedback from massive stars at low metallicities: MUSE observations of N44 and N180 in the Large Magellanic Cloud}

\author[A. F. McLeod et al.]{A. F. McLeod,$^{1,2,3}$\thanks{E-mail: anna.mcleod@canterbury.ac.nz},
J. E. Dale$^{4}$,
C. J. Evans$^{5}$,
A. Ginsburg$^{6}$,\newauthor
J. M. D. Kruijssen$^{7}$,
E. W. Pellegrini$^{7}$,
S. K. Ramsay$^{8}$
and L. Testi$^{8,9}$
\\
$^{1}$School of Physical and Chemical Sciences, University of Canterbury, Science Road, Christchurch, New Zealand\\
$^{2}$Department of Astronomy, University of California Berkeley, Berkeley, CA 94720, USA\\
$^{3}$Physics Department, Texas Tech University, PO Box 41051, Lubbock, TX 79409, USA\\
$^{4}$Centre for Astrophysics Research, University of Hertfordshire, College Lane, Hatfield, AL10 9AB, UK\\
$^{5}$UK Astronomy Technology Centre, Royal Observatory Edinburgh, Edinburgh, UK\\
$^{6}$National Radio Astronomy Observatory, Socorro, NM 87801 USA\\
$^{7}$Astronomisches Rechen-Institut, Zentrum f{\"u}r Astronomie der Universit{\"a}t Heidelberg, Heidelberg, Germany\\
$^{8}$European Southern Observatory, Karl-Schwarzschild-Str. 2, 85748 Garching bei M{\"u}nchen, Germany \\
$^{9}$INAF/Osservatorio Astrofisico of Arcetri, Largo E. Fermi, 5, 50125 Firenze, Italy
}

\date{Accepted XXX. Received YYY; in original form ZZZ}

\pubyear{2015}

\begin{document}
\label{firstpage}
\pagerange{\pageref{firstpage}--\pageref{lastpage}}
\maketitle

\begin{abstract}
We present MUSE integral field data of two {\sc HII} region complexes in the Large Magellanic Cloud (LMC), N44 and N180. Both regions consist of a main superbubble and a number of smaller, more compact HII regions that formed on the edge of the superbubble. For a total of 11 HII regions, we systematically analyse the radiative and mechanical feedback from the massive O-type stars on the surrounding gas. We exploit the integral field property of the data and the coverage of the He{\sc II$\lambda$}5412 line to identify and classify the feedback-driving massive stars, and from the estimated spectral types and luminosity classes we determine the stellar radiative output in terms of the ionising photon flux $Q_{\mathrm{0}}$. We characterise the HII regions in terms of their sizes, morphologies, ionisation structure, luminosity and kinematics, and derive oxygen abundances via emission line ratios. We analyse the role of different stellar feedback mechanisms for each region by measuring the direct radiation pressure, the pressure of the ionised gas, and the pressure of the shock-heated winds. We find that stellar winds and ionised gas are the main drivers of HII region expansion in our sample, while the direct radiation pressure is up to three orders of magnitude lower than the other terms. We relate the total pressure to the star formation rate per unit area, $\Sigma_{\mathrm{SFR}}$, for each region and find that stellar feedback has a negative effect on star formation, and sets an upper limit to $\Sigma_{\mathrm{SFR}}$ as a function of increasing pressure.

\end{abstract}


\section{Introduction}
Throughout their lifetime, massive stars influence their environment by injecting energy and momentum into the surrounding interstellar medium (ISM) via a series of feedback mechanisms, e.g. protostellar outflows, stellar winds, ionising radiation, supernovae (see \citealt{krumholz14} for a review). On large, kpc scales, feedback from massive stars dominates the mass and energy balance in star-forming galaxies like the Milky Way and plays a crucial role in the evolution of galaxies by regulating the formation of new generations of stars (e.g. \citealt{ostriker10}). On smaller, $<$100 pc scales, massive stars are responsible for inflating HII regions, heating and ionising the surrounding gas, and adding to the turbulent velocity field of the local ISM \citep{krumholz16}. Qualitatively, the effect of massive stars on their environment is well understood, but a solid quantitative analysis is still missing. As a result, feedback from massive stars represents a key uncertainty in galaxy formation studies, as simulations can only reproduce the observed galaxy population by tuning the strength of stellar feedback (\citealt{schaye15}, \citealt{vogel14}).

Giant HII regions and super-bubbles around massive stars and star clusters are among the most prominent features of star-forming galaxies. Their evolution links stellar feedback from the small to large scales, as they are believed to represent evolved HII regions, the expansion of which is driven by stellar winds and supernovae \cite{oey962}. However, to understand and evaluate the role stellar feedback plays in driving the dynamics of these regions, it is imperative to characterise the massive stellar content within them. The past two decades have seen many studies tackle feedback, ionisation structures, morphologies, dynamics and stellar content of feedback-driven bubbles and HII regions (e.g. \citealt{lopez14}, \citealt{lopez11}, \citealt{pellegrini11}, \citealt{pellegrini10}, \citealt{oey96dyn}, \citealt{oey96star}, among many others), in an ever-increasing effort aimed at disentangling and observationally quantifying the impact of the various feedback mechanisms. 

Optical integral field spectroscopy (IFS) is ideally suited to analyse both the gaseous (\citealt{mcleod16}, \citealt{kehrig16}, \citealt{m16}, \citealt{monreal11}) and stellar (\citealt{castro18}, Zeidler et al. subm.) components of Galactic and nearby extragalactic HII regions. The advantage of IFS over conventional imaging or spectroscopy is the fact that it combines both, which allows the simultaneous derivation of kinematics and physical quantities of the feedback-driving stars and feedback-driven gas with the same data set. The power of IFS data to link the massive stellar population to feedback-driven structures in HII regions has been demonstrated in \cite{mcleod16b}, where we showed a tight correlation between the rate at which molecular cloud structures are photo-evaporated by stellar feedback, and the amount of ionising photon flux originating from the nearby O-type stars.

In this work, we exploit optical integral field data from the MUSE instrument mounted on the Very Large Telescope (VLT), to analyse the properties and kinematics of the ionised gas in two HII region complexes in the Large Magellanic Cloud (LMC), and connect these to the populations of massive, feedback-driving O-type stars that have formed in the two regions. The LMC is an ideal target for stellar feedback studies, as it is nearly face on (hence allowing a convenient viewing angle); it hosts numerous massive star-forming regions; it does not suffer from the heavy extinction we would face if observing Galactic HII regions; at 50 kpc, it is at an ideal distance to observe entire regions within a feasible amount of telescope time and yet still be able to resolve and classify individual stars. Furthermore, the LMC is characterised by a lower metallicity ($\sim$ 0.5 solar), a lower dust-to-gas ratio \citep{roman14}, and a corresponding environment in which the effect of ionisation is enhanced \citep{israel86} and the impact of radiation pressure subdued, therefore allowing one to probe stellar feedback in an lower-metallicity environment. 

As part of a pilot program to map HII regions in the Magellanic Clouds with MUSE we obtained 8' x 8' mosaics of the giant HII regions N44 (DEM L152) and N180 (DEM L326). Both regions consist of a central, giant bubble, on the rim of which smaller and more compact HII regions have formed. Because of the large number of targeted HII regions of different sizes and morphologies, these two targets offer a superb opportunity of investigating stellar feedback for a multitude of stellar contents and bubble parameters with just two data sets. N44 is one of the most luminous HII regions in the LMC after 30 Doradus, it contains 3 OB associations (LH 47, 48 and 49, \citealt{lucke70}), and it is associated with a supernova remnant located about 6 arcminutes NE of the main central bubble. N44 seems to be more massive and more evolved than N180: \cite{oey95} classify $>$30 O-type stars, and find evidence for two episodes of star formation, an earlier ($\sim$10 Myr) episode which occurred inside the main bubble, and a later ($\sim$5 Myr) episode outside of the latter. N180 hosts the young ($<$10 Myr, \citealt{bica96}) cluster LH 117 \citep{lucke70} which contains several O-type stars, as well as over 30 young stellar objects \citep{caulet08}.

This paper is organised as follows: Section \ref{obs} describes the observations, data reduction and calibration; in Section \ref{stars_sec} we proceed in spectroscopically and photometrically identifying and classifying the massive stellar content of N44 and N180; following this, we characterise the two HII regions in terms of their luminosities, morphologies, sizes, densities, ionisation structure and kinematics in Section \ref{charac}; in Section \ref{fb} we combine the information about the feedback-driving massive stars and the feedback-driven gas to analyse the radiative and mechanical feedback, and discuss photon leakage in Section \ref{photons}; finally, we discuss (Section \ref{discussion}) and summarise (Section \ref{summary}) our findings.

\section{Observations}\label{obs}
\begin{figure*} 
\centering
\mbox{
\subfloat[]{\includegraphics[scale=0.39]{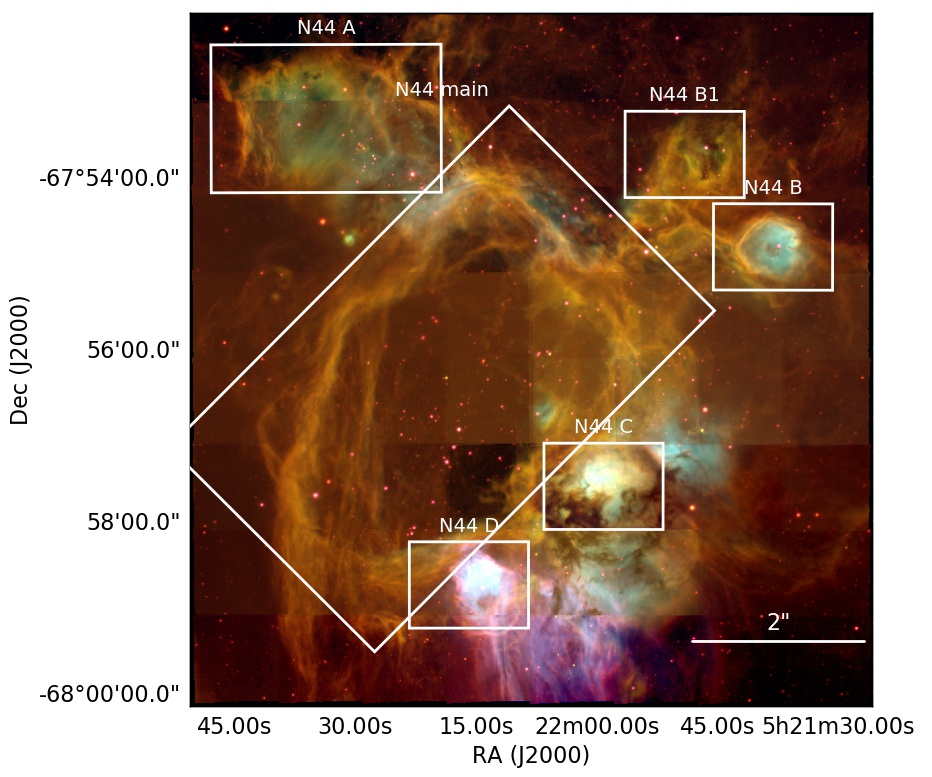}}
\subfloat[]{\includegraphics[scale=0.39]{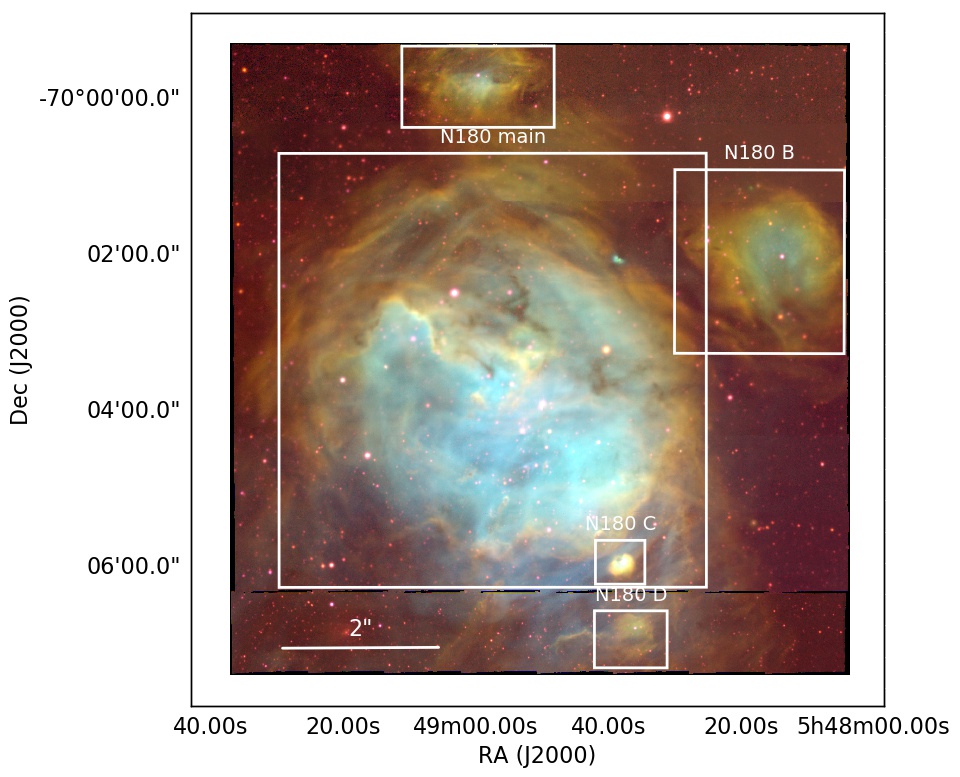}}}
\caption{Three-colour composites of N44 (a) and N180 (b). Red corresponds to [SII]$\lambda$6717, green to H$\alpha$, and blue to [OIII]$\lambda$5007 (stretched logarithmically). Here and in all following figures north is up and east is left. Suboptimal observing conditions during the observations of N44 lead to lower spatial resolution in three observing blocks. The sub-regions analysed in this work are marked in white.}
\label{rgb}
\end{figure*}

All observations were taken with the integral field spectrograph MUSE \citep{muse} mounted on the VLT under the program 096.C-0137(A) (PI McLeod). In the Wide Field Mode used for this program, MUSE has a pixel scale of 0.2 arcsec/pixel, a resolving power of 1770 to 3590 (from 4750 to 9350 \AA\ ), and a spectral separation of 1.25 \AA\ between single frames. Both HII regions were observed with a 64-pointing mosaic, spanning a total size of 8 x 8 arcminutes. Each pointing was observed twice in a 90$^{\circ}$ rotation dither pattern with an integration time of 90 seconds per exposure. Each mosaic therefore consists of 128 single telescope pointings. The observations were scheduled into 1-hour observing blocks of 16 pointings each. Hence, the mosaics are divided into 8 horizontal stripes of 8 final data cubes, where each of the stripes corresponds to one observing block. Due to the filler nature of this program, different observing blocks were executed on different dates with widely varying observing conditions. The consequence of a lower seeing-limited spatial resolution due to the poor observing conditions is still noticeable as can be see in Fig.~\ref{rgb}, where the lower-quality observing blocks results in the stripy nature of the image. We will discuss the effect of this on the analysis in the following sections.

The data was reduced using the MUSE pipeline \citep{pipeline} under the \textsc{Esorex} environment and the standard static calibration files which come with the installed pipeline. For each observing block we used the available calibration files from the ESO archive for the relevant night. The resulting two data cubes of each pointing (one per 90$^{\circ}$ rotation telescope pointing) were then combined into a final data cube by using the built-in exposure alignment and combination package of the MUSE pipeline. 

We proceeded in the same manner as with our previous MUSE programs and analyses (\citealt{m16,mcleod16,mcleod16b}) in producing extinction-corrected maps of the main nebular emission lines, velocity maps, and electron density maps. We exploit the wavelength coverage of MUSE to compute extinction maps via the Balmer decrement for the entire observed mosaic of both regions. For this, we use \textsc{pyneb} \citep{pyneb} and assume an extinction curve for the LMC as in \citet{gordon03}. Fig.~\ref{rgb} shows three-colour composites obtained from the extinction-corrected [SII]$\lambda$6717 (red), H$\alpha$ (green) and [OIII]$\lambda$5007 emission line maps. To highlight both the morphology of the gas and the location of the feedback-driving stars, the maps in Fig. \ref{rgb} are not continuum-corrected. For the nebular emission line maps of interest in this paper (H$\alpha$, H$\beta$, [SII]$\lambda$67171,31, [NII]$\lambda$6584, [OIII]$\lambda$5007) we produce continuum maps from a portion of the spectrum close to the relevant lines which does not show any kind of line structure (nebular emission lines or sky lines), by using the same spectral width used to compute the emission line maps (namely $\pm$3 \AA\ around the lines of interest, as summarised in Table~\ref{continuum}). We subsequently subtract the continuum maps from the emission line maps. We note that the continuum subtraction of this emission line map leads to negative values where stellar emission is over-subtracted. In the extinction correction procedure, negative values are set to nan values, and because the dereddening is based on the H$\beta$ map, the pixels with nan values are propagated to the other emission line maps. This leads to apparent stellar emission at the locations of the stars in the electron density map (see Section \ref{lum}), where nan values are interpolated over during the smoothing procedure.

\begin{table}
\centering
\small
\begin{tabular}{lc}
\hline
\hline
Emission line & $\lambda_{\mathrm{cen}}$ (continuum) \\
 & \AA\ \\
\hline
H$\beta$ & 4945 \\
OIII5007 & 5035 \\
H$\alpha$, [NII]6584 & 6617 \\
SII6717,31 & 6760 \\
OII7320,30 & 7210 \\
\hline
\hline
\end{tabular}
\caption{Central wavelengths of the spectral extraction (width = 6 \AA\ see text Section 2) used to produce the continuum maps from the MUSE data cubes.}
\label{continuum}
\end{table}

To test the performance of the data reduction and flux calibration we compare the MUSE data to archival H$\alpha$ narrow-band data from the SHASSA\footnote{Southern H$\alpha$ Sky Survey Atlas, \citealt{gaustad01}.}. Comparing MUSE fluxes to other narrow-band optical data requires mimicking said narrow-band filters with the MUSE cube. Hence, we make an emission line map from the MUSE data by summing the flux in the frames contained in the filter specifications of the SHASSA data, corresponding to a 32 \AA~ wide filter centred on the H$\alpha$ line. We then perform aperture photometry\footnote{For this we use the Photutils package, http://www.astropy.org} on the MUSE and SHASSA maps on circular apertures with a 3.85' radius centred on the two regions (5:48:49.46 -70:03:19.53 and  5:22:08.13 -67:56:08.26 for N180 and N44 respectively). We estimate errors on the aperture fluxes by providing an error map to the aperture photometry routine, obtained by assuming a generous 25\% error on the flux of the emission line maps. Table \ref{muse_shassa} shows the result of the comparison: the two datasets agree within 5\% of flux values, and we are therefore confident of the reduced data products.

\begin{table}
\begin{center}
\caption{Comparison of the H$\alpha$ fluxes obtained from the MUSE data set presented in this work, and from archival SHASSA data \citep{gaustad01}. Fluxes are obtained by performing aperture photometry on circular regions and are in units of 10$^{-10}$ erg cm$^{-1}$ s$^{-2}$, and are pre-extinction correction.}
\begin{tabular}{cc}
\hline 
\hline
\multicolumn{2}{c}{N44} 
\\
F$_{\mathrm{MUSE}}$ & F$_{\mathrm{SHASSA}}$ \\
\hline
6.99$\pm$(6$\times$10$^{-4}$) & 7.21$\pm$0.10 \\
\hline
\hline
\multicolumn{2}{c}{N180}
\\
F$_{\mathrm{MUSE}}$ & F$_{\mathrm{SHASSA}}$ \\
\hline
4.64$\pm$(5$\times$10$^{-4}$) & 4.50$\pm$0.06 \\
\hline
\hline
\label{muse_shassa}
\end{tabular}
\end{center}
\end{table}

\section{Characterisation of the massive stellar content}\label{stars_sec}
To compare the {\sc HII} regions in terms of their morphologies, kinematics, ionisation diagnostics and feedback characteristics described in the following sections, it is necessary to first understand and characterise their massive stellar content. We exploit the MUSE IFS data to identify and classify the massive O-type stars based on a method consisting of the following steps\footnote{A public release of the automated O-star detection and classification from MUSE data is planned for the near future, which will be followed by an automated general spectroscopic classification algorithm. To obtain more information, please contact the lead author. }:

\begin{enumerate}
\item spectra of the brightest ($m_{\mathrm{V}}<21$) objects are extracted from all of the 64 individual MUSE cubes available per region;
\item the spectra of the sources identified in (i) are cropped to a relevant portion around He{\sc II}$\lambda$5411 line ($\lambda_{\mathrm{min}}=5350$ {\AA},  $\lambda_{\mathrm{max}}=5500$ {\AA}), the continuum is fitted and subtracted, and the level of noise $N$ is determined as the standard deviation of the continuum;
\item stars are flagged as O-type if within $\sim$$\pm$15 {\AA} of the He{\sc II} line (5400 - 5430 {\AA}) we find at least 2 data points larger than 3.5$N$ to determine the presence of the absorption line; 3.5$N$ was chosen as a reference value, given that a lower limit results in a large number of false positive detections, while a higher limit yields a large number of non-detections;
\item the spectra flagged as O-type are then inspected by eye to check the performance and validity of the He{\sc II} line identification routine;
\item the equivalent width (EW) of the He{\sc II}$\lambda$5411 and He{\sc I}$\lambda$4922 absorption lines in the spectra of the visually confirmed O-stars are computed; here, we fit the absorption lines with Gaussian, Lorentzian and Voigt profiles, and use the best fitting model based on $\chi^{2}$-minimisation;
\item the ratio of the EWs of the two lines (R = EW$_{4922}/$EW$_{5411}$) is then used to determine the spectral type according to the empirical relations given in \citet{kerton99};
\item finally, we perform aperture photometry on the collapsed (i.e.~summed) portion of the cube covering the Johnson-Cousins V band ($\sim$ 4950 - 6050 {\AA}), and with a distance of 50 kpc to the LMC and a mean extinction towards the LMC of A$_{\mathrm{V}}\sim$ 0.38 mag \citep{imara07}, we obtain absolute magnitudes $M_{\mathrm{v}}$ for the O stars, and infer the luminosity class of the stars according to \citet{martins05}.
\end{enumerate}

The source identification in step (i) is performed with dendrograms\footnote{Astrodendro (http://www.dendrograms.org/)}. The dendrogram pruning settings are chosen such that faint ($m_{\mathrm{V}}>21$) sources are automatically ignored. This does not lead to any losses in terms of unidentified O-type stars (at the distance of the LMC this magnitude cut corresponds to an absolute magnitude of $M_{\mathrm{V}}=2.5$, where late-type O stars have $M_{\mathrm{V}}\sim-3.9$ and brighter). Finally, for each star we determine the ionising photon flux $Q_{\mathrm{0,\star}}$ (in photons/s) according to the calibrations given in \citet{martins05}. For the purpose of this work we limit our analysis to O-type stars, which dominate in terms of ionising photon flux. N44 and N180 have 24 and 6 known B-type stars \citep{skiff14}, respectively, and with an indicative flux of log($Q$) $\sim$ 48 s$^{-1}$, this amounts to only about 10\% and 5\% of the respective integrated values for the O star population. For the magnitudes derived as per step (vii), we find mean errors of $\pm$0.019 and $\pm$0.016 for all stars in N44 and N180, respectively.

The classification method used here comes with a series of potential caveats, which we now briefly discuss. One assumption was that of the assumed distance and extinction on which the estimate of the luminosity class is based. Most of our stars have published $(B-V)$ colours in the references listed in Tables 3 and 4. These are typically $-$0.1-0.2\,mag, so consistent with our assumed extinction above. In a couple of cases (LH\,117-98 and LH\,117-144) $B-V$\, $\sim$ 0.00\,mag, implying an extinction of $\sim$1\,mag (assuming $B-V_0$ = $-$-0.3 for an O-type star). LH47-182 is the furthest outlier, with $B-V$\,$=$\,0.13 mag, resulting in $A_V$\,$\sim$1.5\,mag. Compared to the calibrations from \citet{martins05} these three stars should still be robust to within one luminosity class, sufficient for our study of stellar feedback. A follow-up study of the entire stellar populations of N44 and N180 (down to the observational limit) with a classification based on stellar parameters is planned for the near future. Next, line infilling from strong nebular emission could lead to classifications towards earlier spectral types, and hence an overestimate of $Q_{\mathrm{0,\star}}$. We do not expect this to be a major issue for N44 and N180, as He{\sc I}$\lambda$4922 is prone to nebular contamination only for exceptionally strong nebular emission. Similarly, wind contamination of the He{\sc II}$\lambda$5411 is only problematic in the case of extreme supergiants, which we do not observe here. Finally, we note that to be confidently distinguished from early-type B stars, O9.5 and O9.7 stars ideally need additional diagnostics (e.g. SiIII and NII lines).

The identified O-type stars are listed in Tables \ref{n44stars} and \ref{n180stars}, together with their coordinates, their derived spectral types and inferred luminosity classes, apparent magnitudes, photon fluxes and radial velocities (determined from the HeII lines). For the last two we list values obtained from this work, and where available, we also list the spectral type found in literature. Our sample appears to be complete in terms of correctly identifying stars already classified in previous studies. We also identified 10 previously unknown O-type stars (five per region). In the following two subsections we will describe the O-star identification process and the classified objects for N44 and N180 separately.

\subsection{O-type stars in N44}

In terms of number of O stars, N44 is the more populous of the two analysed regions. Past studies (\citealt{will97}, \citealt{oey95}, \citealt{conti86}, \citealt{rousseau78}) have already classified a total of 34 O stars within the region covered by our MUSE observations. As already mentioned in the introduction, in addition to having formed more stars, N44 appears to be slightly older than N180 when it comes to the stellar population. Furthermore, it hosts the evolved Wolf-Rayet star WN4b \citep{hainich14}, located in N44 main (see Fig. \ref{rgb} concerning the nomenclature of the various subregions). For this class of Wolf-Rayet stars, \citet{crowther07} give a Lyman continuum ionising flux of log(Q$_{HI}$) $\sim$ 49.2 s$^{-1}$.

From the MUSE data we are able to correctly identify 33 of the 34 literature stars, while LH 47-14, classified as an O9.5 V star in \citet{oey95}, has been reclassified as a class B star in this work based on both the HeII and HeI EWs (0.29 and 0.79, respectively). Out of the identified O stars, 10 clearly lie within the bubble that is N44 main (see Fig.~\ref{stars}a). N44 A in the NE corner of the FOV contains 3 O stars, two known and a newly classified one. N44 B, B1 and D all contain a single O star each, out of which the O5 V source in N44 D is the most luminous one (correspondingly, N44 D is one of the brightest regions in N44). N44 C hosts 3 O stars, among which LH 47-191 is the most luminous in the entire observed region. The remaining O stars are scattered around the main bubble and the smaller surrounding HII regions, some do not appear to be associated with nebulosities. Where these concentrate in (at least projected) groups, smaller ionisation fronts are identified (e.g. just NW of N44 C, and at the top of N44 main). The summed photon flux of all O-type stars in N44 amounts to log(Q$_{0,\star}$) $\sim$ 50.40 s$^{-1}$ (including the WR star). Despite being one of the brightest regions in the LMC, this is an order of magnitude less than the flux emitted by R136 in 30 Doradus \citep{crowther98}, and almost 4 times weaker than the flux in NGC 3603 (one of the most massive star-forming regions in the Milky Way, \citealt{mcleod16b}). Spectra of all identified stars are shown in Appendix \ref{n44spec}.

\begin{sidewaystable*}
\vspace{-18cm}
\hspace{-1cm}
\footnotesize
\begin{tabular}{lccllcclccccl}
\hline
\hline
Star & R. A. & Dec. & Spectral Type & Reference & EW$_{\mathrm{HeII}}$$^{a}$\tnotex{tnote:a} & EW$_{\mathrm{HeI}}$$^{a}$\tnotex{tnote:a} & Spectral Type$^{b}$\tnotex{tnote:b}  & $m_{\mathrm{V}}$ & $m_{\mathrm{V,MUSE}}$ & log(Q$_{0,\star}$) & v$_{\mathrm{r}}$ & Notes  \\
 & (J2000) & (J2000) & literature &  & &  & (this work) & & & s$^{-1}$ & km s$^{-1}$ & \\
\hline
LH 48-9 & 05 22 37.13 & -67 53 22.8 & O7 III & \citet{conti86} & 0.89 & 0.28 & O7 III & 13.50 & 13.46 & 49.13 & 306.82 & N44 A\\
LH 48-65 & 05 22 13.48 & -67 53 38.2 & O9 III & \citet{oey95} & 0.49 & 0.35 & O8.5 I & 12.92 & 12.81 & 49.19 & 300.07 & \\
LH 48-4 &  05 21 54.95 & -67 53 53.5 & O9 V & \citet{oey95} & 0.48 & 0.62 & O9.5 III & 14.24 & 14.01 & 48.42 & 299.85 &\\
Sk-67 83 & 05 21 50.44 & -67 53 13.6 & B0.5 III & \citet{massey95} & 0.47 & 0.55 & O9.5 III & 13.10 & 13.07 & 48.42 & 300.25 & N44 B1\\
Sk-67 82 & 05 21 46.76 & -67 53 38.1 & B0.5 & \citet{rousseau78} & 0.59 & 0.62 & O9 III & 13.20 & 13.07 & 48.65 & 321.16 &\\
LH 48-111 & 05 22 21.61 & -67 54 08.9 & O5 V((f$^{*}$)) & \citet{oey95} & 1.17 & - & O5 V & 14.43  & 14.41 & 49.22 & 308.69 & weak HeI, N44 A\\
LH 48-33 & 05 22 04.19 & -67 54 28.1 & O8 V & \citet{oey95} & 0.57 & 0.47 & O8.5 V & 14.23 & 14.15 & 48.27 & 260.34 &\\
LH 48-14 & 05 21 58.44 & -67 54 28.0 & O8.5 V & \citet{oey95} & 0.63 & 0.53 &O8.5 V & 14.13 & 14.13 & 48.27 & 330.50 & \\
LH 48-28 & 05 22 02.71 & -67 54 22.8 & O9 V & \citet{oey95} & 0.53 & 0.57 & O9 V & 14.20 & 14.21 & 48.06 & 278.94 &\\
LH 48-26 & 05 22 01.86 & -67 54 16.6 & O9.5 V & \citet{oey95} & 0.60 & 0.43 & O9.5 V & 14.09 & 14.08 & 47.88 & 286.69 & \\
$\big[$WBD97$\big]$ N44 6 & 05 21 37.65 & -67 54 48.5 & O8 III & \citet{will97} & 0.88 & 0.42 & O8 III & 13.40 & 13.37 & 48.88 & 295.72 & N44 B\\
LH 47-275 & 05 22 07.31 & -67 56 04.2 & O8.5 V & \citet{oey95} & 0.51 & 0.43 & O8.5 V & 13.88 & (14.15) & 48.27 & 311.69 & main\\
$\big[$WBD97$\big]$ N44 3 & 05 22 03.79 & -67 55 28.8 & B0.5 Iab & \citet{will97} & 0.24 & 0.61 & O9.5 I & 12.60 & (13.97) & 49.00 & 306.76 & main, use HeI only\\
LH 47-47 & 05 21 47.34 & -67 55 40.8 & O9.5 V & \citet{oey95} & 0.20 & 0.58 & O9.5 V &  13.49 & (13.97) & 47.88 & 264.78 & clear HeII\\
LH 47-407 & 05 22 20.57 & -67 56 48.3 & O9.5 V & \citet{oey95} & 0.28 & 0.64 & O9.5 V & 13.84 & (14.25) &  47.88 & 309.30 & main\\
LH 47-320 & 05 22 11.87 & -67 56 41.1 & O9.5 V & \citet{oey95} & 0.41 & 0.58 & O9.5 V & 14.47 & (15.07) &  47.88 & 295.14 & main\\
$\big[$WBD97$\big]$ N44 7 & 05 22 07.49 & -67 56 45.7 & O9 III & \citet{will97} & 0.77 & 0.21 & O7 III & 13.40 & (14.20) & 49.13 & 321.35 & main\\
$\big[$WBD97$\big]$ N44 4 & 05 21 57.87 & -67 56 25.6 & O7 III & \citet{will97} & 0.67 & 0.19 & O6.5 III & 13.00 & (13.85) & 49.23 & 315.61 &main\\
LH 47-222 & 05 22 02.63 & -67 56 23.9 & O8 V & \citet{oey95} & 1.36 & 0.51 & O7.5 V & 14.71 & (14.97) & 48.61 & 311.36 & main\\
LH 47-138 & 05 21 57.06 & -67 56 59.6 & O8.5 V & \citet{oey95} & 0.73 & 0.53 & O8.5 V & 14.29 & (14.89) & 48.27 & 319.86 & \\
LH 47-66 & 05 21 49.17 & -67 56 50.0 & O9 V & \citet{oey95} & 0.56 & 0.37 & O8 V & 14.65 & (15.29) & 48.44 & 297.13 \\
LH 47-58 & 05 21 48.24 & -67 56 22.4 & O9.5 III & \citet{oey95} & 0.16 & 0.72 & O9.5 III & 14.10 & (14.83) & 48.42 & 312.90 & use HeII only\\
LH 47-433 & 05 22 22.74 & -67 58 05.8 & O9.5 V & \citet{oey95} & 0.47 & 0.56 & O9 V & 14.15 & 14.25 & 48.06 & 306.25 & main\\
LH 47-402 & 05 22 20.39 & -67 57 48.1 & O6.5V((f) & \citet{oey95} & 0.97 & 0.33 & O7 V & 14.41 & 14.44 & 48.75 & 314.33 & main \\
$\big[$WBD97$\big]$ N44 5 & 05 22 17.81 & -67 57 09.2 & O9 III & \citet{will97} & 0.49 & 0.41 & O8.5 III & 13.30 & 13.38 & 48.75 & 290.00 & main\\
LH 47-277 & 05 22 07.49 & -67 58 02.9 & O9.5 V & \citet{oey95} & 0.48 & 0.68 & O9.5 V & 15.04 & 14.93 & 47.88 & 293.27 &main \\
LH 47-191 & 05 21 59.75 & -67 57 36.7 & O5 III(f) & \citet{oey95} & 0.92 & 0.09 & O5 III & 14.14 & 13.49 & 49.48 & 301.30 & N44 C\\
$\big[$WBD97$\big]$ N44 12 & 05 21 59.53 & -67 57 21.1 & O8 V & \citet{will97} & 0.86 & 0.46 & O8 V & 13.90 & 13.85 & 48.44 & 359.42 & N44 C\\
LH 47-131 & 05 21 56.59 & -67 57 33.7 & O9.5 V & \citet{oey95} & - & 0.64 & O9.5 V & 15.23 & 15.09 & 47.88 & 241.05 & weak HeII, N44 C\\
LH 47-71 & 05 21 49.84 & -67 57 07.6 & O6 V((f)) & \citet{oey95} & 1.07 & - & O5 V & 15.34 & 15.53 & 49.22 & 317.70 & no HeI\\
LH 47-15 & 05 21 43.56 & -67 57 08.7 & O9 V & \citet{oey95} & 0.68 & 0.31 & O7.5V & 14.26 & 14.38 & 48.61 & 298.85 & \\
LH 47-338 & 05 22 13.92 & -67 58 37.3 & O7 V((f)) & \citet{oey95} & 1.16 & - & O5 V &14.20 & 13.61 & 49.22 & 274.11 & no HeI, N44 D\\
LH 47-182 & 05 21 59.00 & -67 58 34.8 & O6 V((f)) & \citet{oey95} & 1.20 & 0.17 & O5.5 V & 14.56 & 14.21 & 49.10 & 255.32 & \\
\hline
MUSE N44-1 & 05 21 50.04 & -67 52 24.1& - & - & 0.58 & 0.48 & O8.5 III & - & 13.77 & 48.75 & 311.11 & field?\\
MUSE N44-2 & 05 22 29.93 & -67 54 02.6 & - & - & 0.90 & 0.57 & O8 V & 15.10$^{c}$\tnotex{tnote:c}  & 14.76 & 48.44 & 304.12 & (N44 A)\\
MUSE N44-3 & 05 21 47.90 & -67 54 54.7 & - & - & 0.65 & 0.53 & O8.5 V & 14.45$^{c}$\tnotex{tnote:c}  & 14.35 & 48.27 & 352.68 & \\
MUSE N44-4 & 05 21 53.81 & -67 54 09.5 & - & - & 0.45 & 0.49 & O9 V& 14.41$^{c}$\tnotex{tnote:c}  & - & 48.06 & 307.82 & \\
MUSE N44-5 & 05 22 09.36 & -67 58 31.2 & - & - & 0.70 & 0.90 & O9.5 V & - & 15.23 & 47.88 & 312.94 & \\
\hline
\end{tabular}
\\[1.5pt]
\begin{tablenotes}
      \item\label{tnote:a}$^{a}$Several spectra, especially the HeII lines of the early-type sources, show clear Lorentzian-shaped absorption lines, while others display Gaussian or Voigt profiles. The displayed values are obtained from applying the closest matching distribution, determined via $\chi$-square fitting.
      \item\label{tote:b}$^{b}$ Spectral types from the ratio of the HeI4922/HeII5411 lines \citep{kerton99}. Adopted luminosity classes are inferred on the basis of the estimated absolute magnitudes for each star, cf. the models of \citet{martins05}.
      \item\label{tote:c}$^{c}$ from \citet{zari04}.
\end{tablenotes}
\caption{O stars identified in N44. Unless specified, literature magnitudes are taken from the cited reference. Apparent magnitudes in parenthesis correspond to stars in the two observing blocks with the worst observing conditions. For these, we use the literature $m_{\mathrm{V}}$ value. Radial velocities are determined from the HeII line.}
\label{n44stars}
\end{sidewaystable*}

In Table \ref{n44stars} a number of stars are listed for which the magnitudes derived from the MUSE data are consistently lower than literature values. This is true for all (and only) stars in observing blocks 4 and 5 which suffered from the worst observing conditions. Hence, for these stars, we use the apparent magnitude from literature.

\subsection{O-type stars in N180}\label{n180_stellar}
In the central bubble (N180 main) we expect to find 14 O-type stars (\citealt{massa03}, \citealt{neubig99}, \citealt{massey89}, \citealt{conti86}). Our algorithm correctly extracted and identified 13 of these as O stars, while one of them, LH 117-19, was not flagged as an O star and inspection of the spectrum confirms the absence of the He{\sc II}$\lambda$5411 absorption line. We therefore suggest that LH 117-19 is an early-type B, rather than an O star. The spectra of the 13 identified N180 main stars, cropped around the He{\sc II} and He{\sc I} lines, are shown in Appendix \ref{n44spec}, Fig.~\ref{heii_spec} and Fig.~\ref{hei_spec}, respectively. Fig.~\ref{heii_spec}(a) shows the clear detection of the He{\sc II} line for the brightest O stars in the main bubble, and Fig.~\ref{hei_spec}(a) illustrates the weakness of the He{\sc I} line as a consequence of the early spectral type for the four most luminous stars (LH 117-43A, LH 117-98, Sk-70 115 and LH 117-214). For these four objects, the spectral type was determined with the EW of the He{\sc II} line only, according to Eq. 2 in \citet{kerton99} rather than based on the ratio of the two. The brightest star in N180 is LH 117-214, an O3 V star, located towards the south-western rim of N180 main. We also detect the two previously classified stars south-east of the main bubble, LH 118-165 and LH 118-182. These do not appear to be associated with N180, nor with ionised gas in their vicinity, and their radial velocities suggest them as large-amplitude binaries. We therefore do not include these in subsequent analyses, but list them in Table \ref{n180stars} for completeness.

\begin{sidewaystable*}
\vspace{18cm}
\footnotesize
\begin{tabular}{lccllcclccccl}
\hline
\hline
Star & R. A. & Dec. & Spectral Type & Reference & EW$_{\mathrm{HeII}}$$^{a}$\tnotex{tnote:a} & EW$_{\mathrm{HeI}}$$^{a}$\tnotex{tnote:a} & Spectral Type  & $m_{\mathrm{V}}$ & $m_{\mathrm{V}}$ & log(Q$_{0,\star}$) & v$_{\mathrm{r}}$ & Notes   \\
 & (J2000) & (J2000) & literature &  & &  & (this work) & & (this work) & s$^{-1}$ & km s$^{-1}$ & \\
\hline
LH 117- 43 & 05 49 09.09 & -70 03 05.20 & O6 V & \citet{massey89} & 1.12 & 0.29 & O6.5 V & 14.08 & 14.09 & 48.88 & 238.82 &\\
LH 117- 62 & 05 48 51.63 & -70 03 12.10 & O9 V & \citet{massey89} & 0.46 & 0.55 & O9 V & 14.75 & 14.56 & 48.06 & 212.58 & \\
LH 117- 14 & 05 48 54.62 & -70 02 29.80 & O6.5((f)) & \citet{massey89} & 0.87 & - & O6.5 V$^{b}$\tnotex{tnote:b} & 14.69 & 14.65 & 48.88 & 66.62 & no He{\sc I} \\
LH 117-103 & 05 49 19.51 & -70 03 37.30 & O9.7 & \citet{massey89} & 0.10 & 0.29 & O9.5 III - B0 III & 13.53 & 13.94 & 48.42$^{c}$\tnotex{tnote:c} & 167.77 & weak He{\sc II} \\
LH 117-43A/140 & 05 48 52.67 & -70 04 16.42 & O3 III(f*)-O4(f*) & \citet{massey89}  & 1.09 & - & O5 V$^{b}$\tnotex{tnote:b} & 13.52 & 13.54 & 49.22 & 231.30 & weak He{\sc I}  \\
LH 117- 98 & 05 48 57.64 & -70 03 34.10 & O9 & \citet{massey89} & - & - & - & 14.94 & 14.81 & 48.06 & 213.2 & literature class \\
Sk-70 115 & 05 48 49.65 & -70 03 57.82 & O6.5 Iaf & \citet{massa03} & 0.72 & - & O7.5 I & 12.17 & 12.26 & 49.31 & 278.51 & weak He{\sc I} \\
LH 117-144 & 05 48 30.85 & -70 04 18.50 & O9 & \citet{massey89} & - & 0.60 & O9.5 V - B0 V & 15.23 & 15.05 & 47.88$^{c}$\tnotex{tnote:c} & 294.05 & weak He{\sc II} \\
LH 117-183 & 05 48 54.56 & -70 04 56.70 & O9 & \citet{massey89} & 0.64 & 0.74 & O9.5 V & 14.24 & 14.95 & 47.88 & 287.22 & \\
LH 117- 13 & 05 48 39.03 & -70 04 58.60 & O8 V & \citet{neubig99} & 0.59 & 0.48 & O8.5 V & 15.14 & 14.49 & 48.27 & 225.8 & \\
LH 117-214 & 05 48 48.72 & -70 05 32.20 & O3/4 & \citet{massey89} & 1.33 & - & O3 V & 13.15 & 13.29 & 49.64 & 224.38 & no He{\sc I}  \\ 
LH 117-206 & 05 48 42.02 & -70 05 22.80 & O8 & \citet{massey89} & 0.88 & 0.46 & O8 V & 15.06 & 14.85 & 48.44 & 242.43 &\\ 
LH 117-16 & 05 48 41.22 &-70 04 27.70 & O7 V & \citet{conti86} & 0.62 & 0.44 & O8 V & 14.04$^{d}$\tnotex{tnote:d} & 13.96 & 48.44 & 271.04 & \\
\hline
MUSE N180-1 & 05 48 48.97 & -70 04 45.80  & - & - & 0.49 & 0.66 & O9.5 V & 15.75$^{d}$\tnotex{tnote:d} & 15.36 & 47.88 & 257.59 &\\
MUSE N180-2 & 05 48 59.53 & -69 59 43.80 & - & - & 0.85 & 0.39 & O7.5 V & 14.86$^{d}$\tnotex{tnote:d} & 14.52 & 48.61 & 232.95 & N180 A \\  
MUSE N180-3 & 05 48 49.06 & -70 00 50.85 & - & - & 0.65 & 0.54 & O8.5 V & 14.48 & 14.75$^{d}$\tnotex{tnote:d} & 48.27 & 231.47 &  \\   
MUSE N180-4 & 05 48 13.70 & -70 02 04.31 & - & - & 0.82 & 0.29 & O7 III & 13.92$^{d}$\tnotex{tnote:d} & 13.85 & 49.13 & 241.50 & N180 B \\
MUSE N180-5 & 05 48 37.94 & -70 05 58.20 & - & - & 0.60 & 0.46 & O8.5 V & 15.95$^{d}$\tnotex{tnote:d} & 15.54 & 48.27 & 223.17 & N180 C$^{e}$\tnotex{tnote:e} \\
\hline
LH 118-165$^{f}$\tnotex{tnote:f} & 05 49 30.62 & -70 06 04.10 & O9 & \citet{massey89} & - & 0.66 & O9.5 V & 15.15 & 14.93 & 47.88 & 77.76 & weak HeII \\ 
LH 118-182$^{f}$\tnotex{tnote:f} & 05 49 21.37 & -70 06 29.80 & O9III & \citet{massey89} & 0.43 & 0.60 & O9.5 III & 13.78 & 13.84 & 48.42 & 62.66 &  \\  
\hline
 MUSE N188-6$^{g}$\tnotex{tnote:g} & 05 48 35.83 & -70 06 48.0 & - & - & 0.13 & 0.99 & B0.5 III & 14.79$^{d}$\tnotex{tnote:d} & 14.89 & 48.06 & 263.62 & N180 D \\
\hline
\end{tabular}
\\[1.5pt]
\begin{tablenotes}
      \item\label{tnote:a}$^{a}$Several spectra, especially the HeII lines of the early-type sources, show clear Lorentzian-shaped absorption lines, while others display Gaussian or Voigt profiles. The displayed values are obtained from applying the closest matching distribution, determined via $\chi$-square fitting.
      \item\label{tote:b}$^{b}$ the MUSE data does not cover the NIII and NIV lines necessary to determine spectral peculiarities such as the (f*) or the ((f)) type.
      \item\label{tote:c}$^{c}$ In this work we adopt the O-type classification (rather than the B-type) for this source, given the detection (although weak) of the HeII line and we aim at exploring the upper limit of Q$_{0}$ values.
      \item\label{tote:d}$^{d}$ from \citet{zari04}.
      \item\label{tote:e}$^{e}$ obscured by HII region material.
      \item\label{tote:f}$^{f}$ As noted in Section \ref{n180_stellar}, LH 118-165 and LH 118-182 are most likely not associated with N180, but they are included here for completeness.
      \item\label{tote:g}$^{g}$ B-type star in N180 D, included here for completeness.
\end{tablenotes}
\caption{Same as Table \ref{n44stars} for N180.}
\label{n180stars}
\end{sidewaystable*}

\begin{figure*}
\centering
\mbox{
\subfloat[]{\includegraphics[scale=0.45]{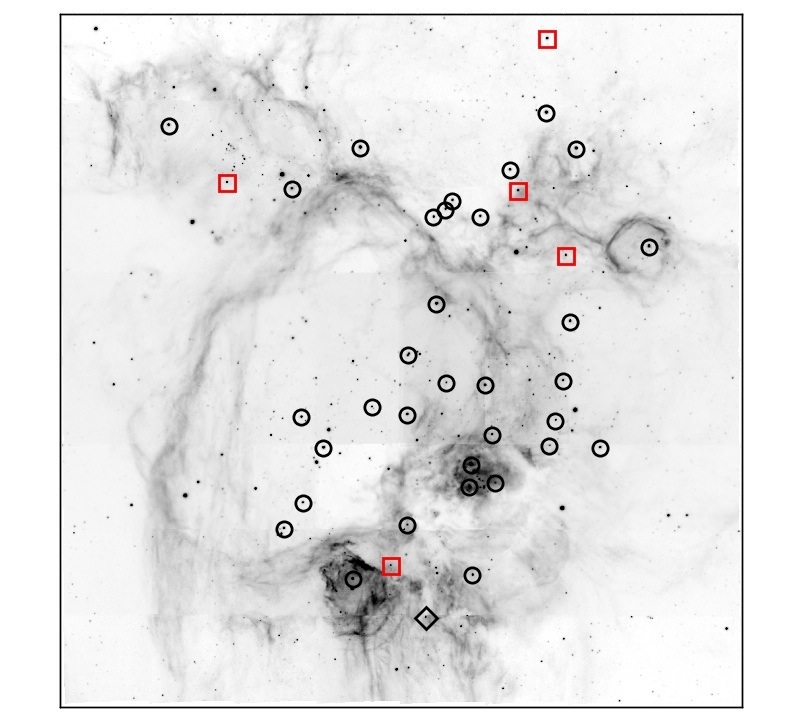}}
\subfloat[]{\includegraphics[scale=0.45]{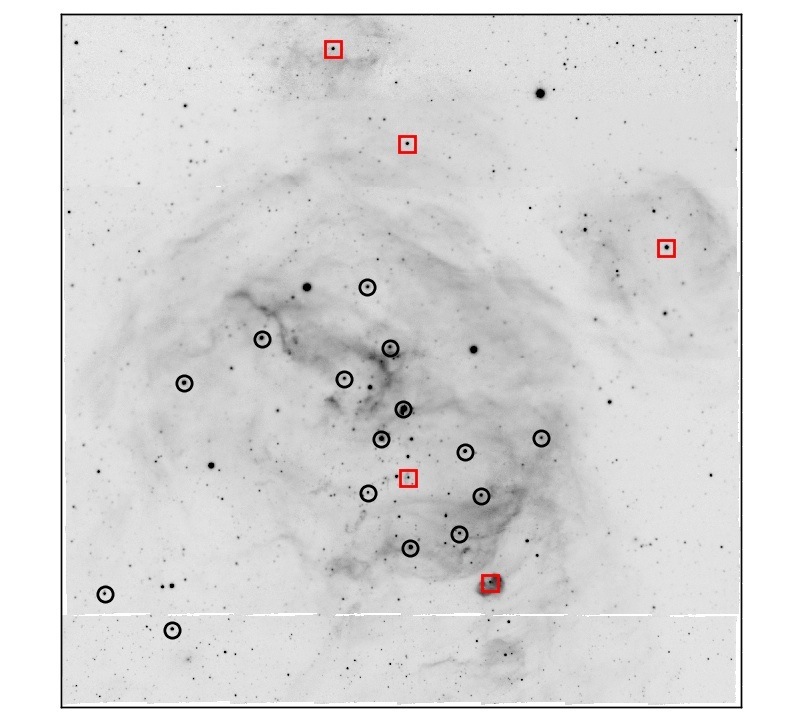}}}
\caption{O-type stars in N44 (a) and N180 (b) on the [SII]6717 map. Black circles mark already known O stars, while red squares mark the objects newly identified in this work. The black diamond in N44 marks the position of the only WR star in the analysed regions.}
\label{stars}
\end{figure*}

In addition to the identification of already known stars, we expand the census of O-type stars in N180 by a total of five newly classified objects (see Fig.~\ref{stars}b). Of these, MUSE N180-1 (a late-type O star) resides within N180 main; MUSE N180-2 is associated with the smaller HII region N180 A north of the main bubble; MUSE N180-4 is associated with N180 B; and finally MUSE N180-5 is driving the small, compact region N180 C immediately south of N180 main.  As discussed in Section \ref{fb}, the surrounding smaller HII regions have radii of the order of 10 to 40 times smaller than the main bubble. The brightest and smallest of these is N180 C, which spans about 6 pc in diameter and contains an O8.5 V star. N180 D contains what is either a late O star or an early B star which displays a very weak HeII line. This star, if of spectral type O, is a false negative of our identification algorithm, as it only has one data point larger than the noise around the line, and thus does not satisfy our selection criterium. Together with an absolute magnitude of $M_{\mathrm{V}}\sim-3.98$, we therefore suggest that this star is an early-type B star rather than a late O star, and assign it a tentative spectral type B0.5 III \citep{wegner06}, and from \cite{sternberg03}, this translates to log(Q$_{0,\star}$) = 48.06 s$^{-1}$. We therefore do not list it in Table~\ref{n180stars}, but given that it appears to be associated with a small HII region where no O star is found, we still include it in the analysis of the following sections. The summed photon flux of all O-type stars identified in the N180 mosaic is log(Q$_{0,\star}$) $\sim$ 50.12 s$^{-1}$.

\section{Characterisation of the HII regions}\label{charac}

\subsection{Sizes and morphologies}\label{radius}
The two regions display different morphologies on about the same physical scales (Fig. \ref{rgb}). Apparent sizes of the various subregions in the two complexes are determined from radial flux profiles, such that the size of a bubble is given by the radius R$_{90}$ which encompasses 90\% of the measured H$\alpha$ flux. The main physical parameters for each region are listed in Table \ref{params}, and Fig.~\ref{n44_profiles} and \ref{n180_profiles} show the radial profiles, together with the limiting radius, for each subregion. 

N44 (DEM L152), located about 1.8$^{\circ}$ NW of 30 Doradus, consists of a superbubble (N44 main, see Fig.~\ref{rgb}) seemingly devoid of ionised gas, along the rim of which several more compact HII regions are located. N44 main is roughly elliptical with a major axis (which has a P.A. of 37$^{\circ}$ west of north) of $\sim$ 65 pc (4.5'), and a minor axis of $\sim$ 44 pc (3'). Except for N44 A, all other compact HII regions have formed along the western rim of N44 main. N44 A is an irregularly-shaped HII region consisting of an eastern bubble around a single O star (LH 48-9), while the majority of the LH 48 cluster stars are located further west, including LH 48-111 and MUSE N44-2. N44 B is a roughly round-shaped HII region around [WBD97] N44 6, with a radius of approximately 6.5 pc. As already noted in \cite{naze02} (where this bubble is referred to as N44 F), two pillar-like structures have formed along the western rim of N44 B. These are protruding into the HII region from the surrounding cloud and are pointing back directly towards [WBD97] N44 6, which is the source most likely shaping these pillars. N44 C is about 14 pc across, it hosts 3 O-type stars and displays dusty filamentary features around its rim. N44 D is among the brightest in N44, and while it is very similar to N44 C in size (R$_{90}\simeq$ 7.4 pc), it hosts a single O5 star (LH 47-338). As can be seen in Fig.~\ref{rgb}, both N44 D and the region surrounding the WR star show bright filaments extending to the SW, and are characterised by very bright [SII] emission. Alongside the morphologically-defined HII regions just described, N44 also displays several regions of ionised gas and associated ionisation fronts, all associated with at least one identified O star. However, given that these do not have a continuous bubble/shell-like structure, we do not include them in our subsequent analyses in this paper.  

N180 (DEM L326), located about 1.3$^{\circ}$ SE of 30 Doradus, consists of a gaseous, round main bubble (N180 main) with a diameter of about 78 pc (5.4'), into which filamentary- and pillar-like structures protrude. At the tip of one of these pillar-like structures an early-type B young stellar object (YSO) of about 12 M$_{\odot}$ \citep{caulet08} has formed, which is launching an 11 pc long bipolar jet. This jet, HH 1177, is the first ionised jet from a high-mass YSO identified outside of our Galaxy, and it is associated with bow-shock structures above and below the pillar where it interacts with the surrounding medium (see \citealt{nature} for a detailed analysis of the object). Four smaller HII regions are located around N180 main, with N180 main hosting the open cluster NGC 2122. All the smaller HII regions around N180 main contain a single O star (or, in the case of N180 D, a B-type star). With a radius of 3.4 pc, N180 C is the smallest of all the subregions. Just south of N180 A we identify a more diffuse region of ionised gas associated with the O7.5 V star MUSE N180-2. 

\subsection{Luminosities, electron densities and ionisation structure}\label{lum}

H$\alpha$ luminosities are measured from circular apertures of radius R$_{90}$ on the MUSE H$\alpha$ map and assuming a distance of 50 kpc to the LMC. We measure the total H$\alpha$ luminosity of the two mosaics to be L(H$\alpha$)$_{\mathrm{N44}}$ $\simeq$ 38.25 erg s$^{-1}$ and L(H$\alpha$)$_{\mathrm{N180}}$ $\simeq$ 38.06 erg s$^{-1}$. We assume a 10\% measurement error on the H$\alpha$ flux from aperture photometry (see Appendix). In the case of N44 the value derived here is about 1.8 times lower than reported in \cite{lopez14} (L14 henceforth), however the region considered by these authors is larger and also encompasses the nebulosity associated with LH 49 about 6 arcminutes south of N44 main (see Fig. 1 in \citealt{naze02}). 
 
\begin{figure}
\centering
\includegraphics[scale=0.55]{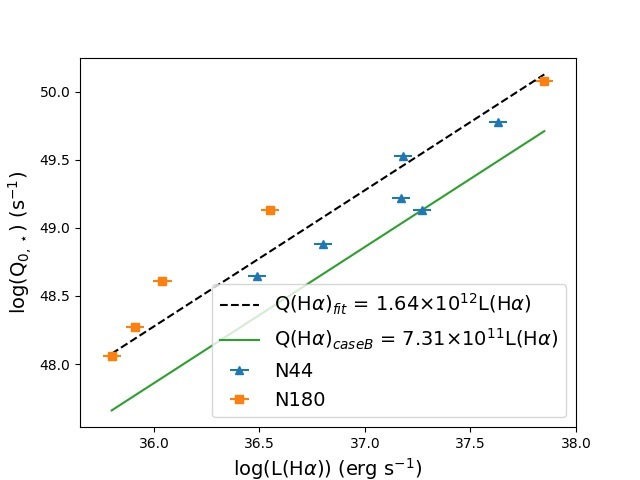}
\caption{Measured H$\alpha$ luminosity vs the ionising photon flux Q$_{0,\star}$ as derived from the spectral types of the stellar population in each region (see Table \ref{pressure}). The dashed black line indicates the best fit to the data, while the solid green line corresponds to the luminosity-photon flux relation as obtained from Case B recombination (\citealt{osterbrock}, see section \ref{lum}). Errors on the measured luminosity are assumed to be $\pm$ 10\% (see Appendix).}
\label{lum_flux}
\end{figure}

With the assumption of Case B recombination (T = 10$^{4}$ K), Q(H$\alpha$)$_{\mathrm{case B}}$ = 7.31$\times 10^{11}$ L(H$\alpha$) s$^{-1}$ (\citealt{osterbrock}), these H$\alpha$ luminosities translate to a number of (log) Lyman continuum photons of 50.10 s$^{-1}$ and 49.91 s$^{-1}$ for N44 and N180, respectively. Luminosities and the computed Q(H$\alpha$)$_{\mathrm{case B}}$ values for the {\it individual} regions are listed in Table \ref{pressure}. We discuss the implications of the difference between the photon flux as derived from the spectral types of the stars in each region and that derived from the H$\alpha$ luminosity in Section \ref{photons}. Here, we test the analytical relation between L(H$\alpha$) and Q(H$\alpha$)$_{\mathrm{case B}}$ by comparing the measured H$\alpha$ luminosity to the photon flux for each region as derived from the stellar content\footnote{In this paper the used notation is as follows:\newline - Q(H$\alpha$)$_{\mathrm{case B}}$ refers to the number of LyC photons as derived according to \citet{osterbrock} for Case B recombination; \newline - Q$_{0,\star}$ refers to the number of ionising photons as measured from the derived spectral types of the analysed stellar population; \newline - Q(H$\alpha$)$_{\mathrm{fit}}$ corresponds to the number of LyC photons as obtained by comparing the measured L(H$\alpha$) to the ionising photon flux as obtained from the stellar population, Q$_{0,\star}$;} (Q$_{0,\star}$). This is shown in Fig.~\ref{lum_flux}, where Q$_{0,\star}$ is plotted against L(H$\alpha$). The linear fit to the data points translates to a relation where

\begin{equation}
Q(H\alpha)_{\mathrm{fit}} = 1.64\times 10^{12} L(H\alpha) \hspace{0.5cm} (s^{-1})
\label{ql}
\end{equation}

When compared to the \citet{osterbrock} Case B relation (hence for an optically thick nebula) between the measured luminosity and the photon flux, Eq. \ref{ql} suggests that the production of an observed H$\alpha$ luminosity via the photon flux injected by a stellar population is less efficient. Hence, an interpretation of this could be that the ratio of Eq. \ref{ql} and Case B corresponds to the fraction of photons that escape the nebula and do therefore not contribute to the luminosity. However, the difference between the Case B relation and the one found here (Q(H$\alpha$)$_{\mathrm{fit}}$) is only about a factor 2, which could be explained with the assumed temperature of 10$^{4}$ K: for a temperature of 7000 K, the difference between Case B and Eq. \ref{ql} is reduced to a factor 1.7 \citep{mckee97} and the conversion of ionising photon flux to H$\alpha$ luminosity is more efficient. Unfortunately, our observations are not deep enough to confidently detect the auroral [NII]$\lambda$5755 line needed to probe temperature via the [NII] emission line ratio. However, based on an electron temperatures of $>10^{4}$ K derived for N44 in \cite{toribio17}, we suggest that temperature is not the key player in explaining the discrepancy. To further strengthen this point we note that another factor which needs to be taken into account when considering HII region temperatures is metallicity. At lower metallicities line cooling is less efficient, and gas temperatures are therefore higher. Conversely, the translation from spectral type to Q$_{0,\star}$ applied here is based on calibrations which assume solar metallicity, which introduces a further uncertainty. We will test the dependance on metallicity with MUSE observations of HII regions in the Small Magellanic Cloud (SMC), and, together with future deeper observations of more LMC (and SMC) HII regions, we will also be able to efficiently probe the electron temperature.   

\begin{figure*}
\centering
\mbox{
\subfloat[]{\includegraphics[scale=0.42]{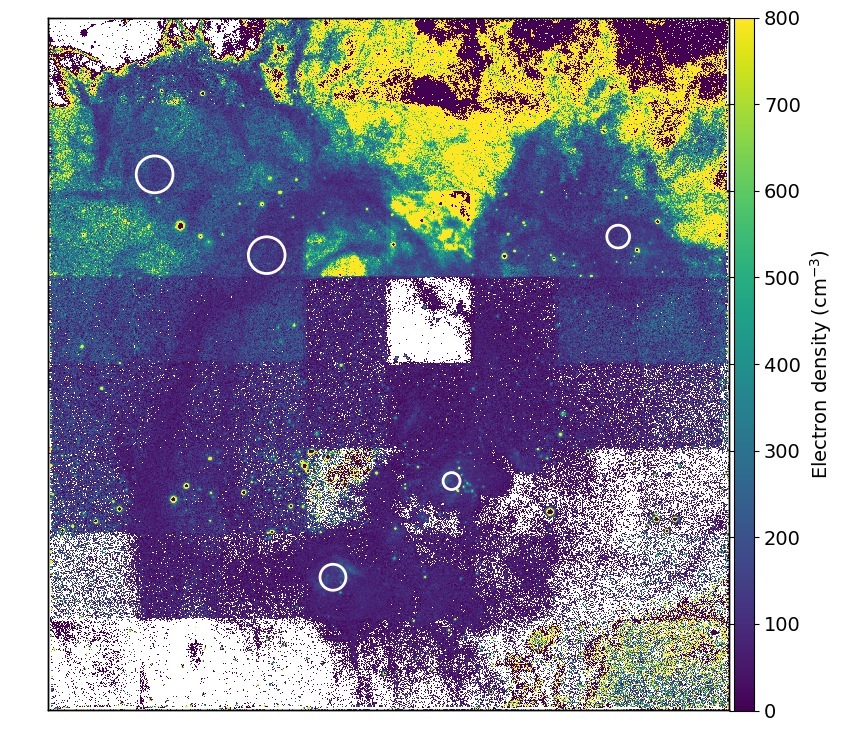}}
\subfloat[]{\includegraphics[scale=0.42]{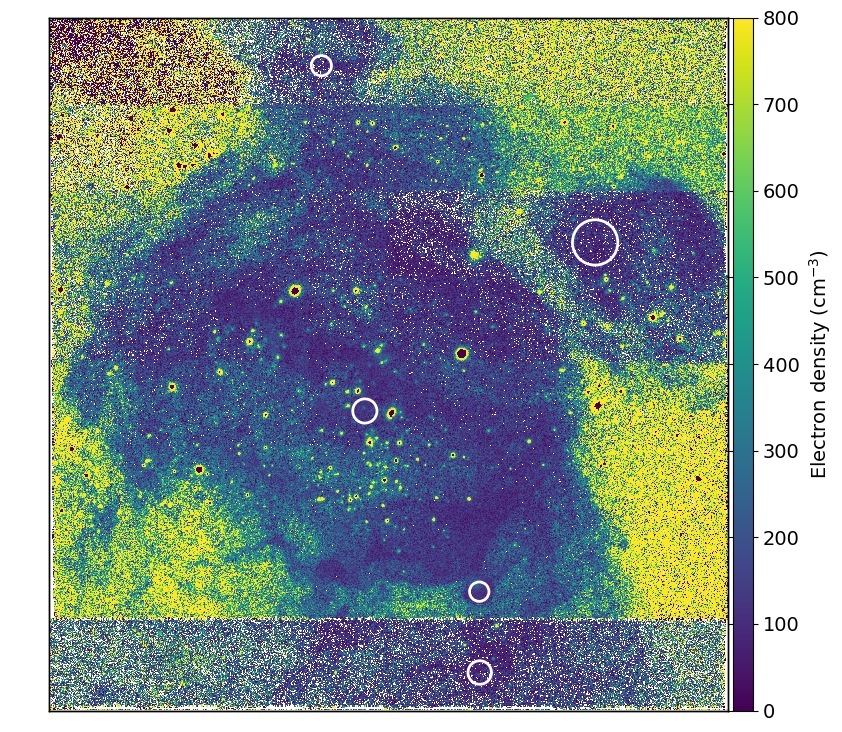}}}
\caption{Electron density ($n_{\mathrm{e}}$) maps of N44 (left) and N180 (right). White circles show the apertures used to extract $n_{\mathrm{e}}$ values reported in Table~\ref{params}. Over-subtraction of the continuum in the H$\beta$ map, used in the dereddening procedure of the emission line maps, leads to residuals at the location of stellar sources (see text Section \ref{obs}.)}
\label{den}
\end{figure*}

We compute the electron density $n_{\mathrm{e}}$ (used later in this paper to determine the pressure of the ionised gas, see \ref{sec_pion}) from the line ratio of the two continuum-subtracted and reddening-corrected sulphur line maps at [SII]$\lambda$6717 and [SII]$\lambda$6731 as in \citealt{m16}, assuming a temperature of 10$^{4}$ K (which is a good assumption for HII regions, \citealt{peimbert17}), and discarding pixel values with [SII]$\lambda$6717/ [SII]$\lambda$6731$>$1.49, as these lead to unphysical negative values (see Eq. 4b in \citealt{mccall84}). The resulting maps (smoothed with a Gaussian kernel) are shown in Fig.~\ref{den}. Despite the mosaic pattern in N44, the morphology of the {\sc HII} regions is recognisable in the density maps. Together with the RGB composites shown in Fig.~\ref{rgb}, this illustrates the lower matter density of the main bubble of N44, the emission from the ionised gas coming from its rim, while the smaller surrounding regions are richer in gas. N180 on the other hand still shows a significant amount of ionised material in the main bubble. 

The two regions show moderate electron densities, with mean values (across the entire mosaics) of $\sim$170 cm$^{-3}$ and $\sim$250 cm$^{-3}$ for N44 and N180, respectively . The values reported here are a factor 2-3 higher than the values reported in L14. These authors compute the electron density from the flux density of the free-free emission at 3.5 cm, such that $n_{\mathrm{e}}$ is proportional to the flux, the square root of the temperature, and inversely proportional to the volume of the emitting region. As noted in \citet{peimbert17}, densities derived from radio emission are generally smaller than those obtained from the ratio of collisionally excited lines (as in this work), and the two are related via the filling factor, i.e. the fraction of a nebula filled with high-density material. Electron densities derived via these two methods are not directly comparable, and we therefore compare the values obtained via the emission line ratio method used here to similar available results from the literature. We note that the apertures used to derive mean electron densities for each region do not contain stellar residuals coming from the continuum subtraction procedure. For N44 we compare the derived mean electron density with that found in \citet{toribio17}, who use VLT-UVES spectra of various LMC (and SMC) HII regions to derive abundances and physical parameters. These authors find $n_{\mathrm{e}}=$ 200$\pm$150 cm$^{-3}$ from the [SII] line ratio for an extraction of 3.0$\times$9.4 arcsec$^{2}$ centred on 05:22:13:6 -67:58:34.2 J2000 (corresponding to N44C, see Fig.~\ref{n44c}). For the same extraction, MUSE yields $n_{\mathrm{e}}=$ 192$\pm$36 cm$^{-3}$, thus in excellent agreement with the UVES results.

\begin{figure*}
\centering
\mbox{
\subfloat[]{\includegraphics[scale=0.43]{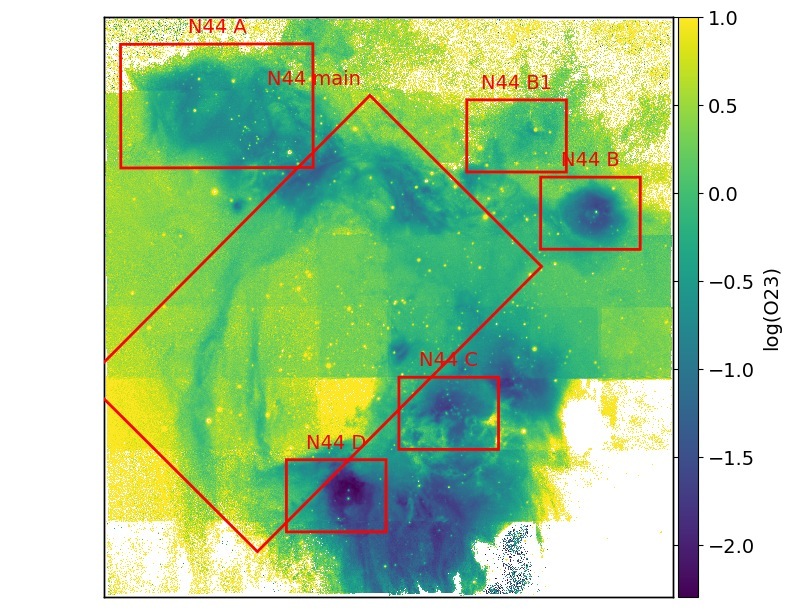}}
\subfloat[]{\includegraphics[scale=0.43]{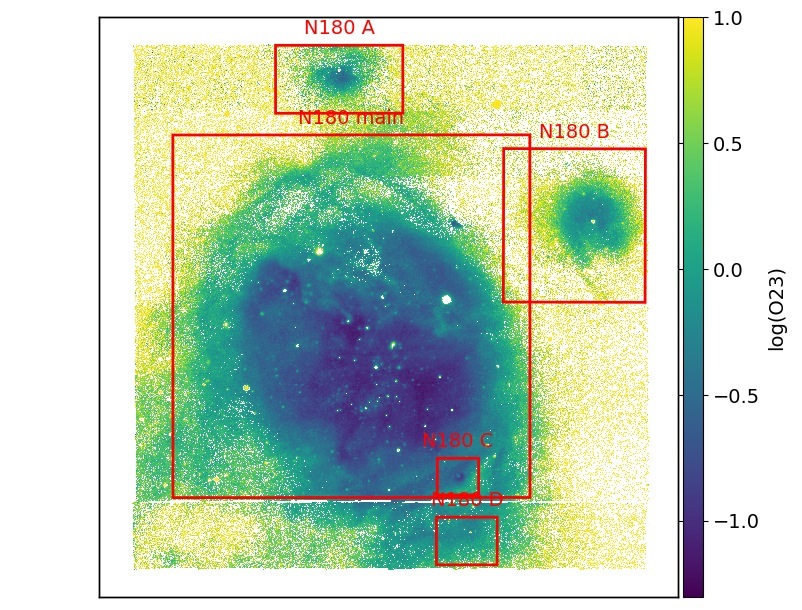}}}
\caption{The emission line ratio O23 (=[OII]$\lambda$7320,7330/[OIII]$\lambda$4959,5007), tracing the degree of ionisation, of N44 (left) and N180 (right). Individual subregions are marked in red to facilitate the discussion in Section \ref{lum}. Stellar residuals can be seen where the continuum subtraction of the [OII] maps is not optimal, due to the selected continuum for these lines being $\sim$100 \AA\ away from the emission lines (see Table \ref{continuum}).}
\label{o23}
\end{figure*}

We investigate the ionisation structure of the two HII regions with the ratio O23 (=[OII]$\lambda$7320,7330/[OIII]$\lambda$4959,5007), which traces the degree of ionisation. This is shown in Fig.~\ref{o23}. In N44, the highest degree of ionisation (lowest O23 values) correspond to the subregion N44 D (N44 A, B and C clearly being traced by higher degrees of ionisation as well), while the low-density cavity of N44 main shows a significantly lower degree of ionisation. The opposite is the case for N180, where the main cavity shows the lowest O23 values, together with N180 D. 

Overall, N180 shows higher O23 values, i.e. a lower degree of ionisation, than N44. With the two regions having similar densities, we suggest that the reason for this is in fact the higher number of ionising sources present in N44 (in combination with their spectral type), with the exception of the evacuated inner parts of N44 main.

\subsection{Gas-phase metallicities}\label{metall}
The so-called {\it strong-line method} (\citealt{alloin79}, \citealt{pagel79}) to determine chemical abundances in Galactic and extragalactic {\sc HII} regions is a widely accepted and used procedure. The method relies on the measurement of emission line ratios, which are then converted into metallicity according to empirically-derived relations that link the line ratios and the metallicity and which are based on oxygen abundances computed via the {\it direct $T_{\mathrm{e}}$-method} (e.g. \citealt{lin17}, \citealt{perez17}, \citealt{pilyugin16}, \citealt{mcleod16}, \citealt{westmoquette13}, \citealt{monreal12}, \citealt{bresolin12}). Despite the uncertainties introduced by the empirical relations, this method of deriving gas-phase metallicities is preferred over theoretical calibrations, as these do generally not reproduce observed oxygen abundances \citep{pilyugin12}.

As MUSE covers all the necessary emission lines to derive the electron temperature $T_{\mathrm{e}}$ (e.g. [NII]$\lambda$5755, [NII]$\lambda$6548, [NII]$\lambda$6584), it would be possible to derive oxygen abundances via the $T_{\mathrm{e}}$-method from this data set. However, given the short exposure time used for this program and due to the [NII]$\lambda$5755 auroral line being very faint and hence indistinguishable from the continuum on a single pixel basis (as already mentioned above), a resolved map of this line would be completely dominated by noise. 
To determine gas-phase metallicities we therefore use the abundance indicators O3N2 and N2 (e.g. \citealt{monreal11}) and compute oxygen abundances according to the empirical relations given in Eq. 2 and 4 in \citet{marino13}. The line ratios used are defined as\footnote{For brevity we omit the wavelength symbol $\lambda$ henceforth when referring to emission lines.}

\begin{equation}
\mathrm{O3N2 = log\bigg(\frac{[OIII]5007}{H\beta}\times\frac{H\alpha}{[NII]6583}\bigg)}
\label{o3n2}
\end{equation}

\begin{equation}
\mathrm{N2 = log\bigg(\frac{[NII]6583}{H\alpha}\bigg)}
\label{n2}
\end{equation}

\begin{figure*}
\centering
\includegraphics[scale=0.8,trim=2cm 0 0 0]{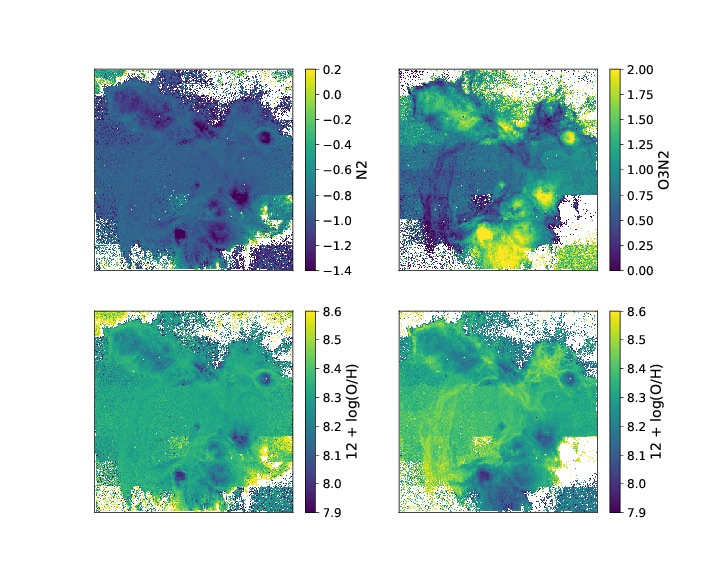}
\caption{The N2 and O3N2 line ratio maps (upper left and upper right, respectively) and the derived oxygen abundances of N44. Over-subtraction of the continuum in the H$\beta$ map, used in the dereddening procedure of the emission line maps, leads to residuals at the location of stellar sources (see text Section \ref{obs}.)}
\label{metal44}
\end{figure*}

\begin{figure*}
\centering
\includegraphics[scale=0.8,trim=2cm 0 0 0]{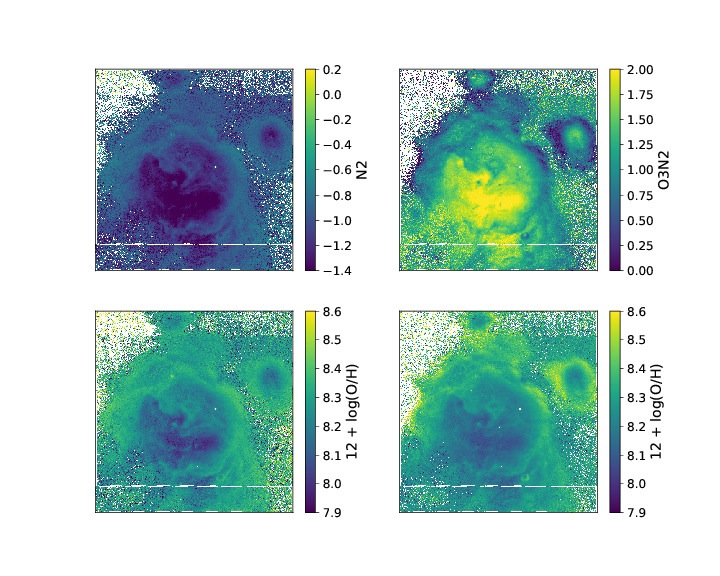}
\caption{The N2 and O3N2 line ratio maps (upper left and upper right, respectively) and the derived oxygen abundances of N180. Over-subtraction of the continuum in the H$\beta$ map, used in the dereddening procedure of the emission line maps, leads to residuals at the location of stellar sources (see text Section \ref{obs}.)}
\label{metal180}
\end{figure*}

Figures \ref{metal44} and \ref{metal180} show the maps of the line ratios given in Eq.~\ref{o3n2} and \ref{n2} (upper panels), as well as the derived oxygen abundances (lower panels). As discussed in e.g. \cite{ercolano12}, deriving abundances of spatially resolved regions from integral field spectrographs introduces a series of caveats. Non-trivial dependencies of the derived values on the density and temperature trace the ionisation structure of the observed nebula, and lead to falsely low abundances for regions with a higher degree of ionisation \citep{mcleod16}. This can be readily seen when comparing the degree of ionisation traced by O23 (Fig.~\ref{o23}) and the derived oxygen abundances (lower panels in Figures \ref{metal44} and \ref{metal180}): low O23 values (high degree of ionisation), found in N180 main and the smaller HII regions, show much lower O/H values than the denser gas that forms the shells. The N2 ratio shows low values at higher degrees of ionisation (given the Balmer line in the denominator), while O3N2 shows a positive relation with O23, due to the dominating [OIII] emission. 

\begin{figure}
\centering
\includegraphics[scale=0.4]{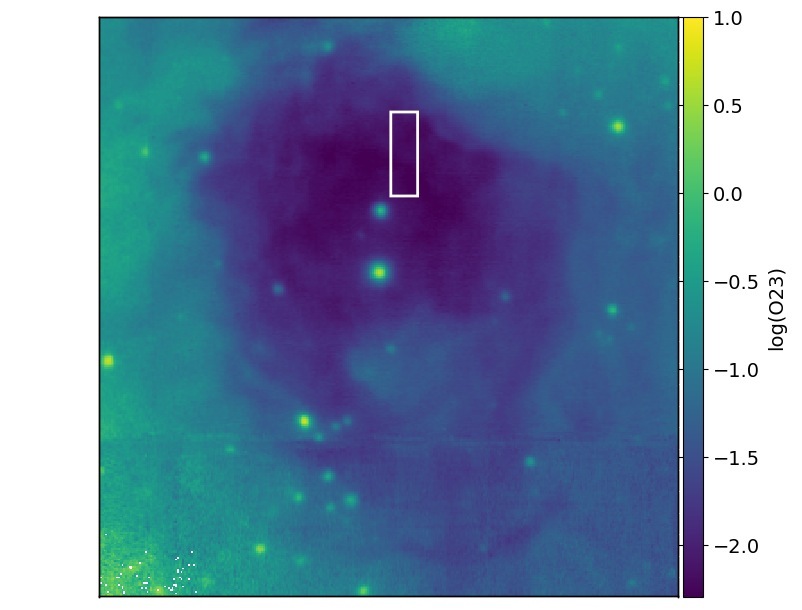}
\caption{ O23 map of N44 D, showing the region analysed in \citet{toribio17} (see text Section 4.3). Stellar residuals can be seen where the continuum subtraction of the [OII] maps is not optimal, due to the selected continuum for these lines being $\sim$100 \AA\ away from the emission lines (see Table \ref{continuum}).}
\label{n44c}
\end{figure}

We compare the obtained O/H values to those derived via the classical $T_{\mathrm{e}}$-method in \citet{toribio17} (henceforth referred to as TSC17). These authors derived ionic and total abundances for LMC and SMC HII regions from VLT-UVES spectroscopic data. Their sample does not include N180, meaning that the following discussion is based on the comparison of values obtained for N44 only. For this region, TSC17 extract a spectrum from a 3.0$\times$9.4 arcsec region centred on 05:22.13.6 -67:58:34.2 (J2000), which is within the small HII region N44 C. A zoomed-in version of the O23 map (Fig.~\ref{n44c}) shows that the extraction region lies within the N44 C HII region (and does not contain stellar residuals resulting form the continuum subtraction, see Section \ref{obs}), therefore displaying a high degree of ionisation. Table \ref{oh_comp} compares the O/H values as obtained via the strong-line method in this work and the $T_{\mathrm{e}}$-method of TSC17. It is clear that the high degree of ionisation in N44 D leads to much lower O/H values when compared to the $T_{\mathrm{e}}$-derived values, and when compared to the mean of all data points.

\begin{figure*}
\centering
\mbox{
\subfloat{\includegraphics[scale=0.4]{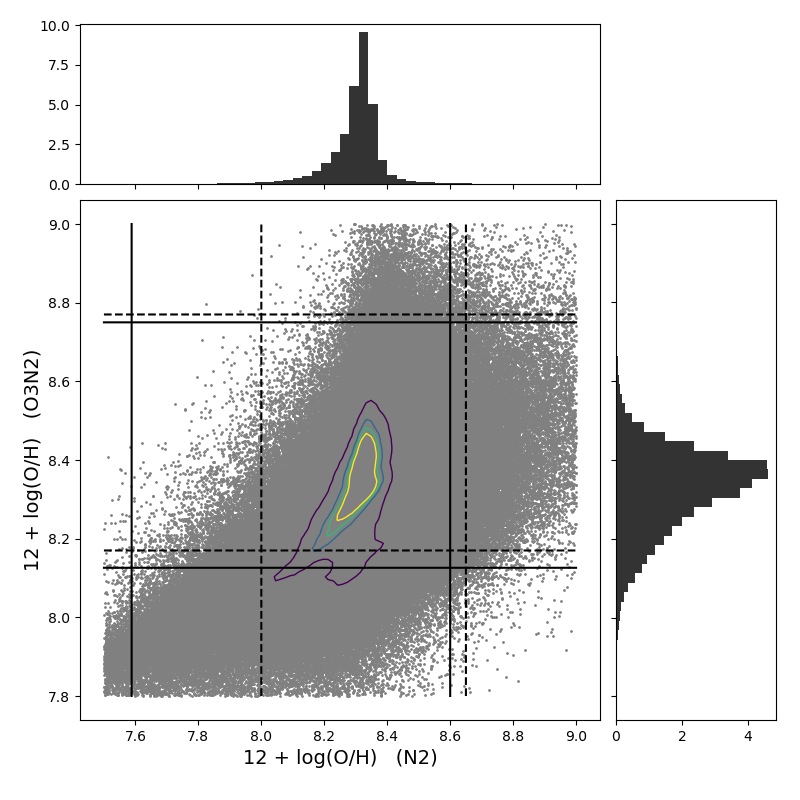}}
\subfloat{\includegraphics[scale=0.4]{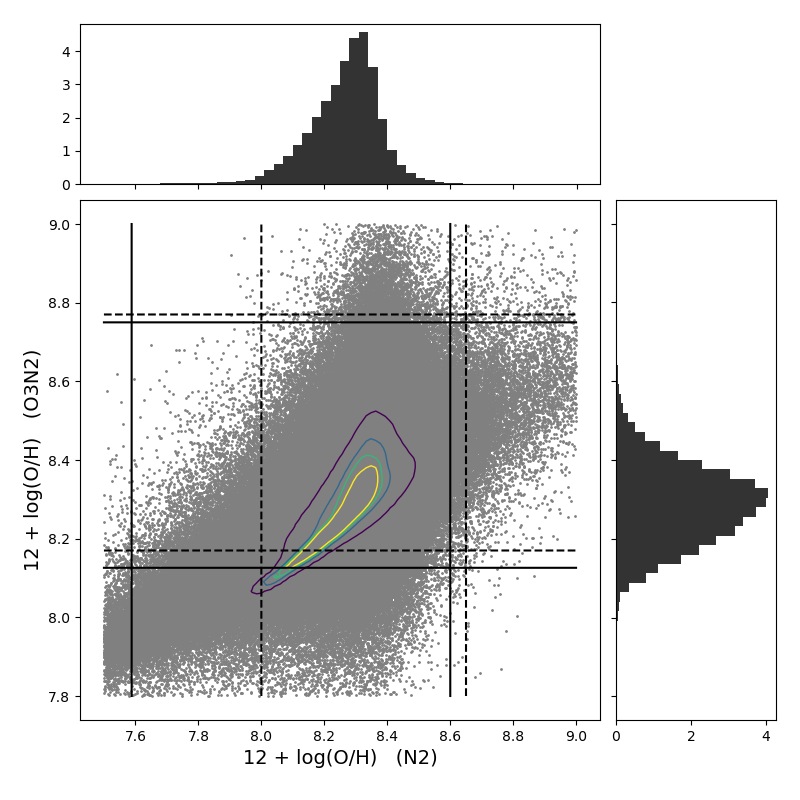}}}
\caption{The oxygen abundance as derived from the N2 line ratio vs. that derived from the O3N2 ratio (left for N44 and right for N180). Contour levels (smallest to largest) correspond to 50\%, 60\%, 75\% and 90\% of data points. Shown are only data points with realistic values, i.e. for which 7.5 $<$ 12 + log(O/H) $<$ 9. Vertical and horizontal lines indicate the applicability (solid lines) and the fit (dashed lines) intervals of the \citet{marino13} calibrations.}
\label{metal_contours}
\end{figure*}

The mean value obtained from all data points reported in Table \ref{oh_comp} agree very well, within errors, with the value from collisionally-excited lines of TSC17. To check for consistency we perform a cut based on a threshold corresponding to the 90\% contour shown in Fig.~\ref{metal_contours}. After the cut we find mean O/H values (from N2 and O3N2, respectively) of 8.30$\pm$0.06 and 8.33$\pm$0.09 for N44; 8.27$\pm$0.10 and 8.28$\pm$0.09 for N180 (stellar residuals, as mentioned in Section \ref{obs}, are masked before deriving mean abundances). For N44, the best agreement with TSC17 is found between the O/H values derived from the collisionally-excited line (CEL) method and the N2 ratio, although the difference between the values obtained from O3N2 and N2 is minimal. The discrepancy between the values obtained in this work and those obtained in TSC17 from recombination lines (RLs) remains large. From a sample of 5 HII regions in the LMC and 4 in the SMC these authors find a better agreement between stellar abundances and those derived from CELs (rather than those derived from RLs) for low-metallicity environments such as the LMC. We therefore take the mean of the oxygen abundances derived from the N2 and O3N2 lines to be a reliable value, which corresponds to 8.32$\pm$0.08 for N44 and 8.28$\pm$0.10 for N180. These agree very well with \citet{russell92}, who find O/H + 12 = 8.35 $\pm$ 0.06 as derived from HII regions in the LMC, and is consistent with an almost flat radial O/H gradient (TSC17). However, further investigation is needed to better understand the dependence of the oxygen abundance not only on the degree of ionisation in spatially-resolved nebulae, but also on metallicity. 

We note that the \citet{marino13} calibrations used here to derive the oxygen abundance were chosen because they are among the calibrations with the most complete HII region observations in terms of metallicities and quality. Other widely used calibrations for N2 and O3N2 are e.g. those of \citet{pettini04}. However, as discussed in \citet{marino13}, the difference between their and the \citet{pettini04} calibrations is mainly caused by the \citet{pettini04} study lacking high-quality auroral line observations. We have tested the difference between the two by replicating the oxygen abundance analysis performed here with the \citet{pettini04} calibrations (not shown here). We find that the steeper O3N2-O/H relation of \citet{pettini04} leads to even lower O/H values for high O3N2 regions, i.e. regions with a high degree of ionisation. By using the \citet{marino13} we still find values lower than the O3N2 fit interval (dashed horizontal lines in Fig. \ref{metal_contours}), however these only contribute to $\sim$ 5\% to the data points in the 90\% contour level (used to determine mean oxygen abundances). Other emission line ratios can be used to derive oxygen metallicities, an excellent example being the R$_{23}$ parameter which does not show a dependence on the ionisation parameter as the ratios used here \citep{relano10}. However, MUSE does not cover the necessary [OII] lines ($\sim$ 3727 \AA) needed in order to confidently use the associated empirical relations to obtain oxygen abundances.

\subsection{Kinematics}\label{kin}
\begin{figure*}
\centering
\mbox{
\subfloat{\includegraphics[scale=0.4]{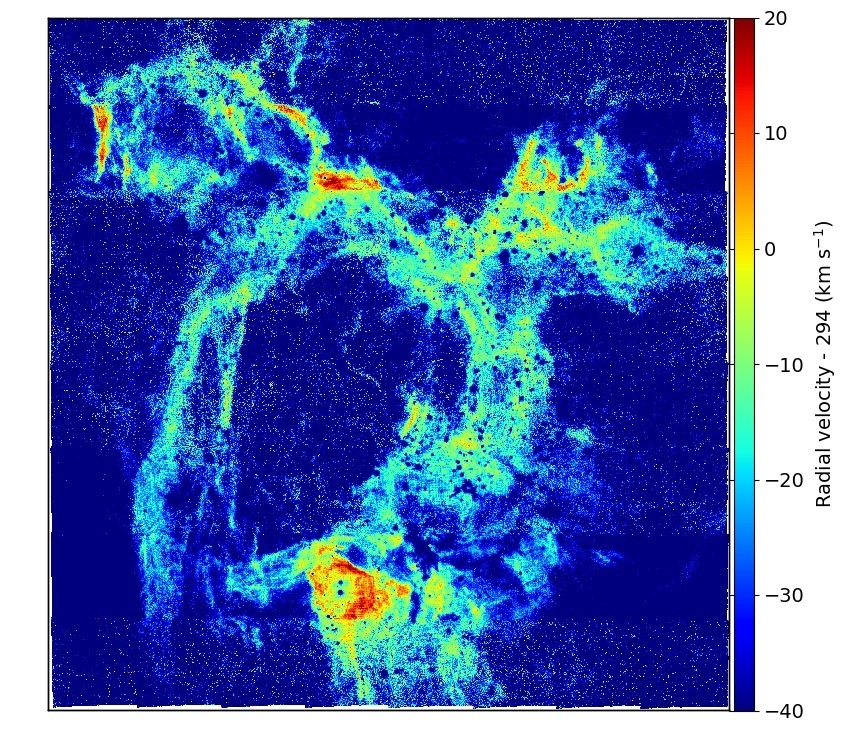}}
\subfloat{\includegraphics[scale=0.4]{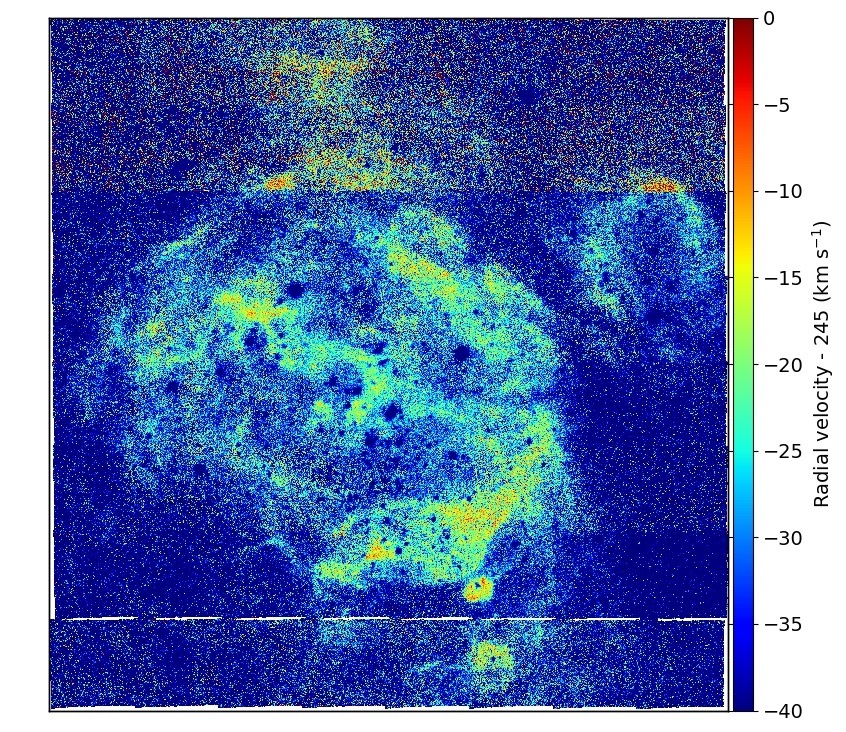}}}
\caption{[SII]6717 radial velocity maps for N44 (left) and N180 (right). }
\label{velmaps}
\end{figure*}

We compute radial velocity maps by fitting emission lines on a pixel-by-pixel basis. For this we proceed as in MC15, where it is shown that stacking a number of different emission lines leads to a better sampling of MUSE spectra. Here, we stack H$\alpha$, [NII]6548,84 and [SII]6717,31, lines originating from the ionised gas. The underlying assumption is that the emission lines used in the stacking procedure are coming from the same line-emitting region. It is therefore a useful procedure to determine radial velocities averaged over entire regions, but in the case of observations of entire spatially resolved HII regions the coverage of a large range of spatial scales with complicated morphologies and sub-structuring, the assumption is no longer valid. Hence, expansion velocities of the individual subregions are derived from the differential velocities across the [SII]6717 velocity maps (shown in Fig. \ref{velmaps}). Given that the [SII] emission is typically more localised than the diffuse H$\alpha$ emission \citep{weilbacher15}, the expanding shells are best traced in [SII].

The shells in N44 are filamentary and sub-structured, and clearly recognisable as expanding ring-like structures. Expansion velocities range from $\sim$ 6 to $\sim$ 11 km s$^{-1}$, in good agreement with literature values (e.g. \citealt{naze02}). The ionised gas in N180 is mainly moving towards the observer, and mean shell velocities are higher than those found for N44. This is supported by \cite{naze01} who find blueshifted expanding shells within N180 main with velocities of 10-20 km s$^{-1}$.

The systemic velocities of the two regions ($\sim$ 294 and 245 km s$^{-1} $ for N44 and N180, respectively) as derived from the ionised gas agree with the global dynamics of the LMC. We demonstrate this by locating the central coordinates of N44 and N180 on the radial velocity map for the LMC as traced by Gaia, as is shown in Fig.~\ref{gaia}. For this, we cross-match the Catalog of Stellar Spectra \citep{skiff14} with the recent Gaia DR2 \citep{gaiadr2}, and select all sources within 10$^{\circ}$ of the LMC. We then separate the sources most likely belonging to the LMC (rather than being foreground sources) by further selecting sources with radial velocities within the range 190-387 km s$^{-1}$ (as determined from HI emission in \citealt{kim98}), and plot these weighted by their radial velocity values. The rotation of the LMC disk is clearly recovered with the Gaia data, as can be seen when comparing the lower panel of Fig. \ref{gaia} with Fig. 3 in \cite{kim98}.

\begin{figure}
\centering
\mbox{
\subfloat{\includegraphics[scale=0.5]{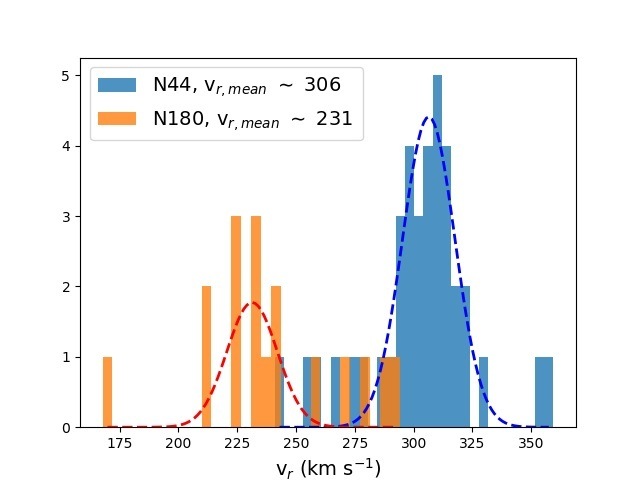}}}
\mbox{
\subfloat{\includegraphics[scale=0.6]{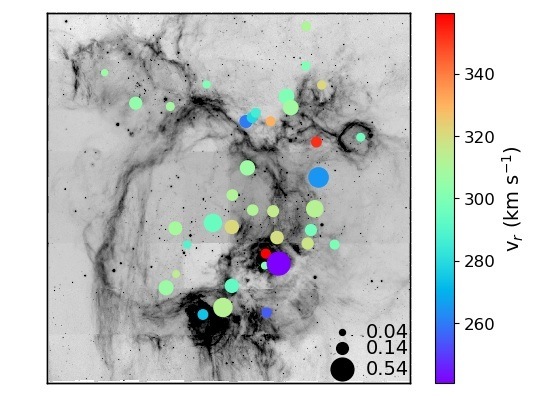}}}
\mbox{
\subfloat{\includegraphics[scale=0.6]{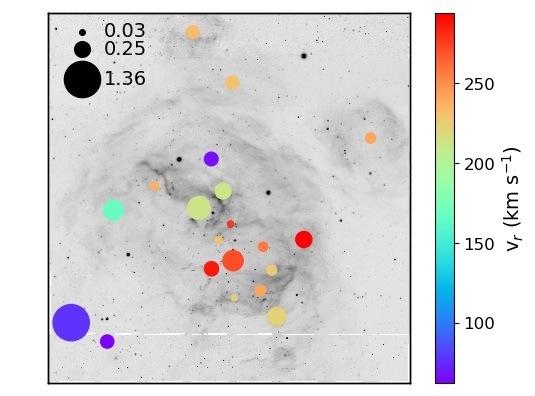}}}
\caption{Upper panel: histograms of the stellar radial velocities, fitted with a Gaussian distribution. Bottom panels: O-type stars colour-coded by radial velocity for N44 and N180 (middle and bottom panels, respectively). The sizes of the circles reflect the errors on the v$_{\mathrm{r}}$ measurement. The two stars with very low radial velocities in N180 (bottom left corner of image) are not included in the histogram fit in the upper panel (see text Section \ref{kin}).}
\label{rv_maps}
\end{figure}

Furthermore, we determine the radial velocities of the O-type stars from the HeII line centroid. The upper panel of Fig. \ref{rv_maps} shows the resulting histograms of the stellar radial velocities, fitted with a Gaussian distribution (for N180, we exclude LH 118-165 and LH 118-182 from the fit, as these are most likely not associated with N180). The lower panels show that there is no obvious relation between the locations of the O-stars and their radial velocity.

\begin{figure}
\centering
\mbox{
\subfloat{\includegraphics[scale=0.5]{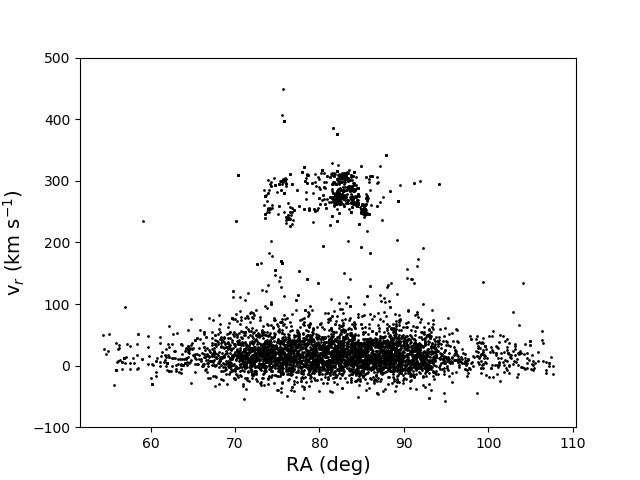}}}
\mbox{
\subfloat{\includegraphics[scale=0.55]{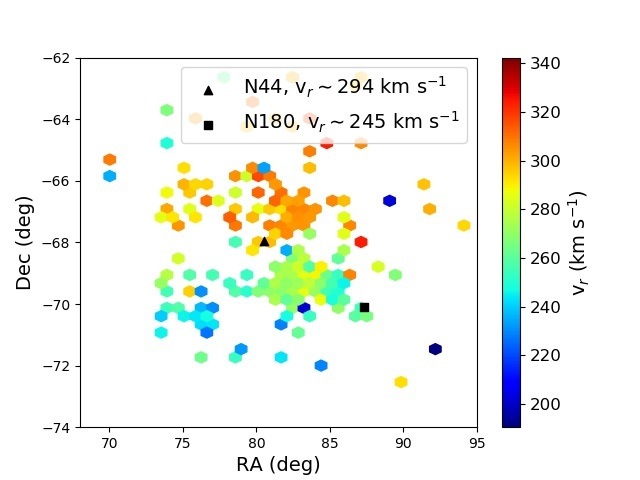}}}
\caption{Upper panel: radial velocity of stars within 10$^{\circ}$ of the LMC, selected by cross-hatching Gaia DR2 and the Catalog of Stellar Spectra \citep{skiff14}. The LMC is distinguishable as the group of data points $>$190 km s$^{-s}$. Lower panel: stars selected from the plot above with 190 $<v_{\mathrm{r}}<$ 387 km s$^{-s}$, the positions of N44 and N180 are indicated with a black triangle and square, respectively.}
\label{gaia}
\end{figure}

\begin{table}
\centering
\small
\begin{tabular}{lccccc}
\hline
\hline
Bubble & Central Coordinates & R$_{90}$ & $n_{\mathrm{e}}$ & v \\
 & (J2000) & (pc) & (cm$^{-3}$) &  (km s$^{-1}$)\\
\hline
N44 main & 5:22:17.2 -67:56:32.8& 45.7 & 133$\pm$43 & 11.5$\pm$6.6 \\
N44 A & 5:22:30.6 -67:53:29.5 & 17.2  & 163$\pm$50 & 10.2$\pm$6.1 \\
N44 B & 5:21:38.3 -67:54:52.2& 6.5 & 129$\pm$31 & 11.1$\pm$5.8 \\
N44 B1 & 5:21:46.4 -67:53:40.9& 9.1 & 132$\pm$32 & 8.4$\pm$6.6 \\
N44 C & 5:21:58.1 -67:57:28.7& 7.2 & 152$\pm$42 & 6.1$\pm$5.4 \\
N44 D &5:22:13.7 -67:58:46.8 & 7.4 & 143$\pm$42 & 9.3$\pm$6.2 \\
\hline
N180 main &5:48:55.2 -70:03:33.5& 39 & 127$\pm$44 & 16.9$\pm$6.7 \\
N180 A &5:49:00.1 -69:59:47.4& 6.6 & 122$\pm$75 & 15.3$\pm$4.7 \\
N180 B &5:48:14.7 -70:02:05.1& 12.2 & 105$\pm$61 & 18.2$\pm$7.3 \\
N180 C &5:48:37.7 -70:06:02.3& 3.4 & 137$\pm$47 & 12.3$\pm$13.7 \\
N180 D &5:48:36.5 -70:06:56.0& 4.6 & 97$\pm$62 & 16.4$\pm$11.7 \\
\hline
\hline
\end{tabular}
\\[1.5pt]
\caption{Central coordinates (column 2), radius containing 90\% of the H$\alpha$ emission (column 3), electron density (column 4) and expansion velocity of the observed subregions. See text Section \ref{charac}.}
\label{params}
\end{table}

\begin{table*}
\begin{center}
\caption{Oxygen abundance in units of 12 + log(O/H) as measured for N44 D (corresponding to the same region as in TSC17, see Fig.~\ref{n44c}) and as measured for the entire mosaic, derived via the two emission line ratios N2 and O3N2, for the mean of all data points, and the $T_{\mathrm{e}}$-derived values reported in TSC17.}
\begin{tabular}{ c c | c c | c c }
\hline 
\hline
\multicolumn{2}{c }{N44 D} 
& 
\multicolumn{2}{c}{N44 all}
&
\multicolumn{2}{c}{\citeauthor{toribio17}}\\
\hline
N2 & O3N2 & N2 & O3N2 & RLs$^{a}$ & CELs$^{b}$ \\
\hline
8.04$\pm$0.06 & 8.02$\pm$0.02 & 8.30$\pm$0.10 & 8.32$\pm$0.11 & 8.58$\pm$0.02 & 8.31$\pm$0.03 \\
\hline
\label{oh_comp}
\end{tabular}
\vspace{1.5pt}
\begin{tablenotes}
      \item\label{tnote:a}$^{a}$Recombination lines.
      \item\label{tnote:b}$^{b}$Collisionally excited lines.
\end{tablenotes}
\end{center}
\end{table*}

\section{Radiative and mechanical feedback}\label{fb}
We now analyse feedback measures related to the radiation and stellar winds emitted by the massive stars that have formed within N44 and N180. These are the direct radiation pressure $P_{\mathrm{dir}}$ (related to the deposition of kinetic energy and momentum by stellar photon flux), the pressure $P_{\mathrm{ion}}$ exerted by the warm (T $ \sim 10^{4}$ K) ionised gas, and the thermal pressure from the shock-heated winds, in a similar fashion as described in L14. These authors derive these and other feedback-related quantities in a comprehensive multi-wavelength analysis of HII regions in the LMC and SMC, ranging from the X-Rays to the infrared, using optical SHASSA data to determine $P_{\mathrm{dir}}$ and $P_{\mathrm{ion}}$. L14 conclude that the pressure of the warm ionised gas dominates over the other pressure terms, with the direct radiation pressure being the weakest term. The radii of the HII regions in their sample range from about 30 to well over 100 pc, and L14 suggest that the role of direct radiation pressure could be better analysed with a sample of younger, smaller HII regions. Except for the two main bubbles, all subregions analysed in this work are an order of magnitude smaller in size than the L14 sample, therefore enabling an extension of the work of L14 towards smaller radii.

\subsection{Direct radiation pressure}
L14 compute the volume-averaged direct radiation pressure assuming a spherical HII region. Because MUSE offers a much higher angular resolution than the SHASSA survey (0.2"/pixel vs. 47.64"/pixel), we can now compute the direct radiation pressure for all subregions rather than averaging over what is clearly a population of different HII regions that vary in age, size and stellar content. For this we use Eq. 3 in \cite{pellegrini07},

\begin{equation}
\begin{aligned}
P_{\mathrm{dir}} = \frac{Q_{0,\star}\langle h\nu \rangle}{4\pi R^{2}c}
\end{aligned}
\label{pdir}
\end{equation}

\noindent where $\langle h\nu \rangle$ is the mean photon energy (here we adopt $\langle h\nu \rangle \sim$ = 15 eV, \citealt{pellegrini07}) and R = R$_{90}$. This form is directly comparable to the radiation pressure force integrated at the ionisation front, where our electron densities and HII regions pressure terms are measured. This differs from the method used in L14, where the direct radiation pressure is derived via the bolometric luminosity measured for an HII region. However, this requires treating P$_{\mathrm{dir}}$ as an energy density and hence assuming full coupling with the gas, which might not be correct for evolved (i.e. evacuated) HII regions. This leads to P$_{\mathrm{dir}}$ as derived via the L14 method being about an order of magnitude higher than what is found here. 

The radiation pressure thus computed for each subregion is listed in Table \ref{pressure}. P$_{\mathrm{dir}}$ is generally of the order of a few 10$^{-13}$ dyn cm$^{-2}$, with the exception of N44 C and N44 D. In addition to the higher P$_{\mathrm{dir}}$ values, these two regions are an order of magnitude more luminous than the other regions of comparable sizes. We will further discuss these two regions in Section \ref{discussion}.

\subsection{Pressure of the ionised gas}\label{sec_pion}

Following L14, the pressure of the ionised gas is computed via

\begin{equation}
P_{\mathrm{ion}}=(n_{\mathrm{e}}+n_{\mathrm{H}}+n_{\mathrm{He}})kT_{\mathrm{e}}\approx2n_{\mathrm{e}}kT_{\mathrm{e}}
\label{pion}
\end{equation}

\noindent which follows from the ideal gas law, under the assumption that helium is singly ionised. As motivated earlier, we assume an HII region temperature of 10$^{4}$ K, and estimate P$_{\mathrm{ion}}$ from the electron density. The pressures of the ionised gas as computed via Eq. \ref{pion} are listed in Table \ref{pressure}. Because of $n_{e}$ being the only free parameter in this pressure term and the subregions all having similar electron densities, the variation in P$_{\mathrm{ion}}$ values is limited, $\sigma$(P$_{\mathrm{ion}}$) $\approx$ 0.5$\times10^{-11}$ dyn cm$^{-2}$. A larger sample of HII regions in different environments is needed to further investigate the ionisation pressure. Together with $T_{\mathrm{e}}$ measurements via the temperature-sensitive [NII] line ratio from deeper MUSE observations, a large sample of observed HII regions would allow to obtain a larger spectrum of P$_{\mathrm{ion}}$ values across a range of different environments, sizes and stellar content.

\subsection{Stellar winds}
At very early, compact stages, the expansion of HII regions is governed by stellar winds. With parsec-scale radii, the HII regions examined in this work do not classify as ultra-compact HII regions for which this would be the case, but given that they present ring-like morphologies and kinematics of expanding shells, and given the existence of wind-blown superbubbles (e.g. \citealt{oey94}), the contribution from winds from O-type stars needs to be addressed. For this we compute the stellar wind luminosity given the observed bubble radii, densities and dynamical timescales, and derive the wind pressure of the hot, wind-shocked gas according to \cite{tielens05}:

\begin{equation}
\begin{aligned}
R&\simeq 32\Big(\frac{L_{w}}{10^{36}\hspace{0.1cm}\mathrm{erg}\hspace{0.1cm}\mathrm{s}^{-1}}\Big)^{1/5}\Big(\frac{0.5\hspace{0.1cm}\mathrm{cm}^{-3}}{\mathrm{n_{0}}}\Big)^{1/5}\\
&\Big(\frac{t}{10^{6}\hspace{0.1cm}\mathrm{yr}}\Big)^{3/5}\hspace{0.5cm}\mathrm{pc}
\end{aligned}
\label{lw}
\end{equation}

\begin{equation}
\begin{aligned}
P_{w}&\simeq 2.3\times10^{-12}\Big(\frac{L_{w}}{10^{36}\hspace{0.1cm}\mathrm{erg}\hspace{0.1cm}s^{-1}}\Big)^{2/5}\Big(\frac{\mathrm{n_{0}}}{0.25\hspace{0.1cm}\mathrm{cm}^{-3}}\Big)^{3/5}\\
&\Big(\frac{10^{6}\hspace{0.1cm}\mathrm{yr}}{t}\Big)^{4/5}\hspace{0.5cm}\mathrm{dyn}\hspace{0.1cm}\mathrm{cm^{-2}}
\end{aligned}
\label{pw}
\end{equation} 

\noindent where L$_{w}$ is the wind luminosity, R = R$_{90}$, $n_{0}$ = $n_{\mathrm{e}}$/0.7 \citep{mcleod16b} and $t$ the dynamical timescale. 

We compare the wind luminosities as computed by rearranging Eq.~\ref{lw} to wind luminosities obtained by directly considering the stellar population in each region, i.e. 

\begin{equation}
L_{w,\star}=\frac{1}{2}\dot M v_{\infty}^{2}
\label{lw_th}
\end{equation}
 
\noindent where $\dot M$ is the stellar wind mass-loss rate and $v_{\infty}$ the terminal wind velocity. For this we use mass-loss rates and terminal velocities from Table 1 in \cite{muijres12} (for $\dot M$ we use the last column, and we compute the terminal velocity as 2.6 times the escape velocity, all used values are listed in Table \ref{lw_star}). As can be seen from Table \ref{pressure}, for seven subregions the wind luminosities derived via Eq. \ref{lw} (which assumes that the bubble size is entirely set by stellar winds) are higher than those predicted from the stellar population. For two subregions, N44 C and N44 D, which also show higher P$_{\mathrm{dir}}$ values, the opposite is the case, while for N44 A and N44 B L$_{\mathrm{w}}$ and L$_{\mathrm{w,\star}}$ are roughly comparable. We will discuss the implications of this in Section \ref{discussion}.

A major uncertainty in the computation of stellar winds from the spectral types of the stellar content is the metallicity used for the prediction of the mass-loss rate. \citet{muijres12} adopt solar metallicity, whereas $\dot M$ scales as $(Z/Z_{\odot})^{\alpha}$, with $\alpha=0.69$ \citep{vink01}. As winds are line-driven, lower metallicities imply fewer lines, hence less momentum transfer, weaker winds and overall less mass-loss. The computed L$_{\mathrm{w}}$ values are therefore almost certainly overestimated and strict upper limits. Dedicated models predicting mass-loss rates and terminal velocities for half-solar metallicity are needed to better compare L$_{\mathrm{w}}$ and L$_{\mathrm{w,\star}}$ in the LMC.

Furthermore, we note that while the theoretical mass-loss rate predictions used to derive wind luminosities from the observed spectral types agree with empirically-derived mass-loss rates (e.g. \citealt{jager88}, \citealt{mokiem07}) in the log(L/L$_{\odot}$) $>$ 5.2 regime, problems occur for O stars with lower luminosities. However, given that it remains unclear whether this problem (referred to as the \textit{weak-wind problem}, e.g. \citealt{marcolino09}) arises from observations or theory and given the good agreement at higher luminosities, for consistency we use the theoretical predictions for the low-luminosity regime as well.

\begin{table*}
\centering
\small
\begin{tabular}{lccccccccccc}
\hline
\hline
Bubble & L(H$\alpha$) & Q(H$\alpha$)$_{\mathrm{case B}}$ & Q$_{0,\star}$ & $f_{\mathrm{esc}}$ & P$_{\mathrm{dir}}$ & P$_{\mathrm{ion}}$ & L$_{\mathrm{w}}$ &  $t_{\mathrm{dyn}}$ & P$_{\mathrm{w}}$ &  L$_{\mathrm{w,\star}}$ & $\Sigma_{\mathrm{SFR}}$\\
 & erg s$^{-1}$ & s$^{-1}$ & s$^{-1}$ & & dyn cm$^{-2}$ & dyn cm$^{-2}$ & erg s$^{-1}$ & Myr &dyn cm$^{-2}$ & erg s$^{-1}$ & 10$^{-6}$ M$_{\odot}$yr$^{-1}$pc$^{-2}$ \\
\hline
N44 main & 37.63 & 49.48 & 49.78 & 0.49 & 0.002  & 3.67 & 37.62 & 3.9 & 1.92 & 36.69 & 0.16\\
N44 A & 37.27 & 49.12 & 49.52 & 0.60 & 0.007 & 4.50 & 36.69 & 1.6 & 1.80 & 36.66 & 0.49 \\
N44 B & 36.80 & 48.64 & 48.88 & 0.42 & 0.012 & 3.56 & 35.84 & 0.6 & 1.67 & 35.81 & 1.18 \\
N44 B1 & 36.49 & 48.34 & 48.65 & 0.51 & 0.004 & 3.64 & 35.79 & 1.1 & 0.98 & 35.38 & 0.29 \\
N44 C & 37.18 & 49.03 & 49.53 & 0.68 & 0.04 & 4.20 & 35.21 & 1.1 & 0.58 & 36.70 & 2.31\\
N44 D & 37.17 & 49.02 & 49.22 & 0.37 & 0.02 & 3.95 & 35.76 & 0.8 & 1.26 & 36.49 &2.13\\
\hline
N180 main & 37.85 & 49.71 & 50.08 & 0.57 & 0.005 & 3.51 & 37.97 & 2.3 & 4.07 & 37.26 & 0.37\\
N180 A & 36.04 & 47.90 & 48.61 & 0.80 & 0.006 & 3.37 & 36.41 & 0.4 & 4.31 & 35.47 & 0.20\\
N180 B & 36.55 & 48.40 & 49.13 & 0.81 & 0.006 & 2.90 & 37.11 & 0.7 & 5.29 & 36.12 & 0.19\\
N180 C & 35.91 & 47.77 & 48.27 & 0.68 & 0.011 & 3.78 & 35.39 & 0.3 & 1.92 & 35.09 & 0.56\\
N180 D & 35.80 & 47.66 & 48.06 & 0.60 & 0.004 & 2.67 & 35.99 & 0.3 & 3.20 & 34.95 & 0.24\\
\hline
\hline
\end{tabular}
\\[1.5pt]
\caption{Column 2: (log) H$\alpha$ luminosity derived from the MUSE H$\alpha$ map. Column 3: (log) number of Lyman continuum photons per second derived via Eq. \ref{oster} from the measured H$\alpha$ luminosity in column 2. Column 4: (log) number of Lyman continuum photons per second emitted by the O-type stars in each region. Column 5: escape fraction (see Eq. \ref{fesc}). Column 6: direct radiation pressure (see Eq. \ref{pdir}). Column 7: pressure of the ionised gas (see Eq. \ref{pion}). Column 8: (log) stellar wind luminosity as derived via Eq. \ref{lw}. Column 9: dynamical timescale obtained from the expansion velocity and radius of each region (Table \ref{params}). Column 10: wind pressure as derived via Eq. \ref{pw}. Column 11: (log) wind luminosity inferred from the spectral type of the stellar content and determined via Eq. \ref{lw_th} using the values given in Table \ref{lw_star}. Column 12: SFR per unit area computed from L(H$\alpha$) via Eq. \ref{sfr}. All pressure terms are in units of 10$^{-10}$ dyn cm$^{-2}$.}
\label{pressure}
\end{table*}

\section{Photon leakage}\label{photons}
Of the ionising photons emitted by massive stars, most are absorbed and therefore go into the heating budget of an HII region. However, a fraction of the photons can escape from the region into the diffuse galactic ISM, hence contributing to the overall energetics of the host galaxy. Determining said fraction of escaping photons is therefore necessary not only to determine the evolution of single HII regions, but also when discussing the impact of massive stars on galaxy evolution. Calculating the escape fraction becomes possible when combining maps of the ionised gas as traced by H$\alpha$ and the knowledge of the stellar content of a given HII region. 

\cite{niederhofer16} use synthetic young clusters to test a method which relies on broadband photometry to determine spectral types of massive stars, stellar atmosphere models to subsequently derive photon fluxes, and then compares the integrated flux from a given stellar population to the same as derived from narrowband H$\alpha$ maps. These authors find that the high uncertainties introduced by determining spectral types from broadband photometry lead to this method not being reliable for determining the escape fraction of ionising photons. Moreover, they suggest that stellar spectra are needed to reduce the spectral type uncertainties, and that such a study could be performed on data of the Magellanic Clouds.

Here, we take up on the results of \citet{niederhofer16} and compare $Q_{\mathrm{0,\star}}$ as obtained by spectroscopically identifying and classifying the O-type stars, to Q(H$\alpha$)$_{\mathrm{case B}}$ \citep{osterbrock} obtained from the measured H$\alpha$ luminosity,

\begin{equation}
Q(H\alpha)_{\mathrm{caseB}}\hspace{0.1cm}(s^{-1})=7.31\times10^{11}L(H\alpha)\hspace{0.5cm}(\mathrm{erg \hspace{0.1cm}s^{-1}})
\label{oster}
\end{equation}

\noindent where L(H$\alpha$) is the H$\alpha$ luminosity of a region as measured from narrowband H$\alpha$ maps, as described in Section \ref{lum}. The fraction of photons escaping from the HII region is then given by 

\begin{equation}
f_{\mathrm{esc}}=\frac{Q_{0,\star}-Q(H\alpha)_{\mathrm{caseB}}}{Q_{0,\star}}
\label{fesc}
\end{equation}

\noindent and the resulting values for each region are listed in Table \ref{pressure}. Escape fractions are $>0.2$ for all regions (consistent for the late stage of HII expansion of low metallicity systems, \citealt{rahner17}), and overall N180 shows higher $f_{\mathrm{esc}}$ values than N44, a fact that is reflected in the distribution of data points in Fig. \ref{lum_flux}. For a statistically significant sample of HII regions in both the SMC and the LMC, \cite{pellegrini12} find that lower column densities lead to leakier HII regions, and that luminous (log(L) $>$ 37 erg s$^{-1}$) regions are predominantly optically thin. Indeed, \citet{pellegrini12} classify N44 as an optically thin region, and N180 as a blister-type region, with column densities of 3.97 and 3.03$\times$10$^{21}$ cm$^{-2}$, respectively. Hence, we suggest that the higher escape fractions found for N180 are a consequence of the lower column density of the complex.

Overall, \citet{pellegrini12} find an escape fraction of $\sim$ 0.42 for the LMC. If the discrepancy between the measured and expected flux-luminosity relation discussed in Section \ref{lum} is indeed tracing the escape fraction, rather than being the consequence of systematic uncertainties, then the resulting $f_{\mathrm{esc}}$ $\sim$ 0.45 is in good agreement with the global value of \citet{pellegrini12} As mentioned before, the preliminary work presented in this paper will be augmented with in-hand/future MUSE observations of over a hundred HII regions in nearby galaxies and the Magellanic Clouds. These will be used to test the dependency of the escape fraction on a variety of HII region parameters such as density, metallicity, radius, etc.

\section{Discussion}\label{discussion}

To discuss the impact of feedback on star formation, we also estimate the star formation rate (SFR) for each subregion according to the relation given in \cite{kennicutt98b},

\begin{equation}
SFR\hspace{0.2cm}(M_{\odot}\hspace{0.1cm}yr^{-1})\hspace{0.1cm}=\hspace{0.1cm}7.9\times10^{-42}\hspace{0.1cm}L(H\alpha)\hspace{0.2cm}(erg\hspace{0.1cm}s^{-1})
\label{sfr}
\end{equation}

\noindent from which (together with R$_{90}$) we derive the SFR per unit area $\Sigma_{\mathrm{SFR}}$, listed in Table \ref{pressure}. We note that the conversion from H$\alpha$ luminosity to SFR in Eq. \ref{sfr} relies on an ensemble average over HII regions of different ages, as well as on population synthesis models. While this introduces caveats when applied to individual objects (as analysed here), it remains a meaningful exercise to estimate the level of star formation activity in our studied regions. 

\begin{figure}
\centering
\includegraphics[scale=0.55]{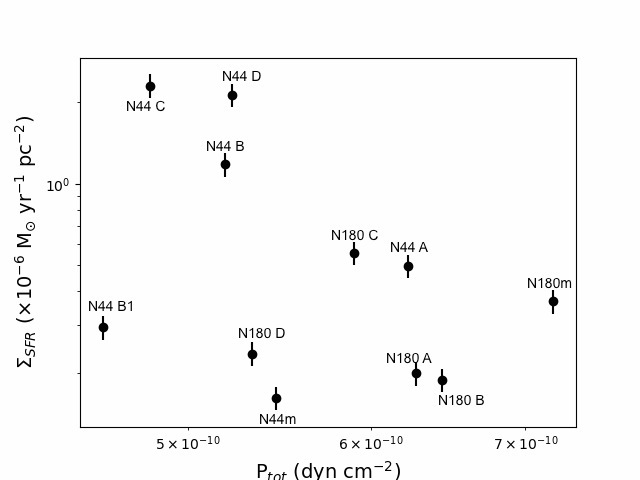}
\caption{The star formation rate (SFR) per unit area as a function of the total pressure for each region.}
\label{ptot_sfr}
\end{figure}

For the total H$\alpha$ luminosities mentioned in Section \ref{lum}, we find total SFRs of $\sim$1400 and 900 M$_{\odot}$ Myr$^{-1}$ for N44 and N180, respectively. The SFR of N44 is thus comparable to that of N11, while both regions have SFRs well below those of 30 Dor or N79 \citep{ochsendorf17}. Fig. \ref{ptot_sfr} shows that together, the analysed feedback mechanisms have a negative effect on star formation by setting an upper limit to $\Sigma_{\mathrm{SFR}}$ as a function of increasing pressure (given by the H$\alpha$ surface brightness cap towards higher feedback pressures). 

\begin{figure}
\centering
\includegraphics[scale=0.55]{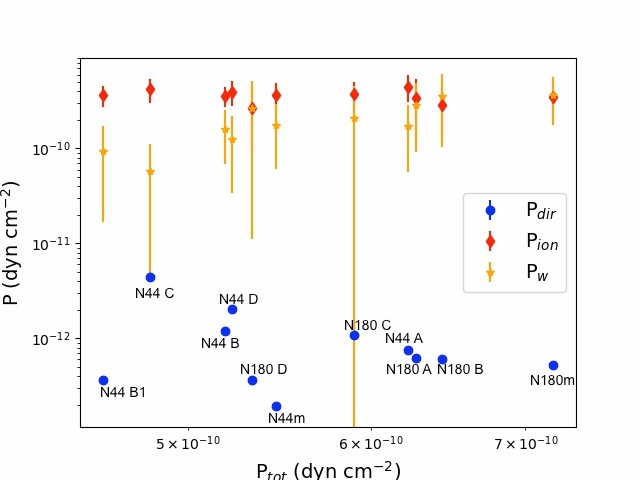}
\caption{Individual pressure terms as a function of total pressure.}
\label{ptot}
\end{figure}

For our sample of smaller, more compact HII regions (compared to the L14 sample) we find that the pressure of the ionised gas generally dominates over the other pressure terms (Fig. \ref{ptot}) and that the direct radiation pressure is the weakest contribution. This is consistent with what L14 find for their sample of HII region complexes. We therefore conclude that none of the HII regions are driven by radiation pressure. As already mentioned in Section \ref{sec_pion}, given the lack of temperature measurements and the similar sizes and electron densities of the N180 and N44 subregions, the MUSE data in this work does not allow a detailed analysis of P$_{\mathrm{ion}}$ as a function of HII region parameters and stellar content. However, while P$_{\mathrm{dir}}$ is about two-three orders of magnitude smaller than P$_{\mathrm{ion}}$, P$_{w}$ is only a factor of a few lower (and in some cases, e.g. N180 B, N180 main, even higher).

When considering the amount of feedback energy injected by stellar winds, we distinguish between three different regimes depending on whether the expected wind luminosity L$_{w,\star}$ (as derived from the spectral types of the stars within the regions) is smaller or larger than the value obtained from a wind-driven model (L$_{w}$ as per Eq. \ref{lw}).

\begin{figure}
\centering
\includegraphics[scale=0.55]{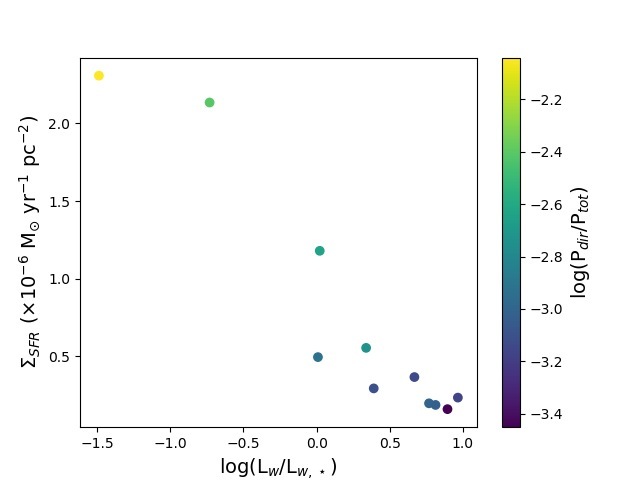}
\caption{The star formation rate per unit areas as a function of the ratio of measured and expected wind luminosity, colour-coded by relative contribution of the direct radiation pressure (P$_{\mathrm{dir}}$/P$_{\mathrm{tot}}$).}
\label{pdir_lw}
\end{figure}

\begin{itemize}
\item \textbf{L$_{w}$ $>$ L$_{w,\star}$}. For seven out the eleven bubbles, wind luminosities as derived via Eq. \ref{lw} are too high with respect to values expected from the stellar content (i.e. derived via Eq. \ref{lw_th}); if the measured radii are correct, then this implies that either the measured expansion velocities are too low, or that stellar winds alone are not enough to account for the measured radii. For these regions we also find the lowest contribution of the P$_{\mathrm{dir}}$ term to the total pressure. 

\item \textbf{L$_{w}$ $\sim$ L$_{w,\star}$}.  For two bubbles, N44 A and N44 B, the wind luminosity as derived from the stellar content (L$_{w,\star}$) is roughly equal to the model value (L$_{w}$, derived via Eq. \ref{lw}), indicating that these are consistent with the standard bubble model \citep{weaver77}.

\item \textbf{L$_{w}$ $<$ L$_{w,\star}$}. The two regions N44 C and N44 D are clear outliers in our sample. With similar sizes as the other subregions, N44 C and N44 D both host early-type O stars. In addition to the O5 III star LH 47-191 (the most luminous in the N44 complex, which alone drives a photon flux of Q$_{0,\star}\sim3\times10^{49}$ photons s$^{-1}$), N44 C also harbours an O8 V and an O9.5 V star. N44 D is being illuminated by LH 47-338, an O5 V star with a photon flux of Q$_{0,\star}\sim1.7\times10^{49}$ photons s$^{-1}$. Both show higher P$_{\mathrm{dir}}$ values, lower P$_{w}$ values, and L$_{\mathrm{w}}$ $<<$ L$_{\mathrm{w,\star}}$. The latter implies a discrepancy between the mechanical energy input from the stellar population in the bubbles and their measured kinematics. In other words, if the expansion velocities are correct, then the bubbles are too small for the amount of energy available from stellar winds alone, as the injected L$_{\mathrm{w,\star}}$ alone would produce bubbles of over twice the observed size of N44 C and D over a period of $\sim$1.5 Myr. \cite{oey96dyn} refers to this as the {\it growth rate discrepancy}, which consists of an apparent overestimate of the wind luminosity found not only for superbubbles, but also for single-star bubbles around e.g. Wolf-Rayet stars (\citealt{oey96dyn} and references therein). Assuming that the measured radii are correct, a possible solution to this would be to increase the expansion velocity by a factor $\sim$ 2-3 to reconcile L$_{\mathrm{w}}$ and L$_{\mathrm{w,\star}}$\footnote{Increasing the expansion velocity would lead to shorter dynamical timescales and therefore to higher L$_{\mathrm{w}}$ values.}. Another possible solution to the discrepancy would be to increase the density by about an order of magnitude. If the measured radii, expansion velocities and densities of N44 C and D are all correct, the question of why these bubbles are not larger still remains. The close proximity of N44 C and D have to the shell of N44 main could for example suggest that the expansion of the two bubbles has been slowed down by high-density material in the direction of N44 main. Indeed CO emission is detected towards N44 C and D, confirming the presence of high-density material associated with these two regions \citep{wong11}. As a consequence of the slower expansion we observe (relatively) higher luminosities, direct radiation pressures and star formation rates for these two regions. This suggests that for small ($<$ 10 pc) HII regions whose expansion is at least partially confined by a high-density environment, a higher photon flux coming from the presence of early-type massive stars (which inject a significant photon flux) leads to an increased contribution of the direct radiation pressure and a decreased contribution of winds towards the total pressure (see Fig. \ref{pdir_lw}). 
\end{itemize}

The three regimes are displayed in Fig. \ref{pdir_lw}, which shows that for L$_{\mathrm{w}}$/L$_{\mathrm{w,\star}}>$ 1, both $\Sigma_{\mathrm{SFR}}$ and the contribution of P$_{\mathrm{dir}}$ to the total pressure are low, while the opposite is the case for the {\it growth rate discrepancy} bubbles (L$_{\mathrm{w}}$/L$_{\mathrm{w,\star}}<$ 1). The three regimes therefore allow a qualitative discussion on the contribution of winds vs. other expansion-driving mechanisms. Indeed, while bubbles for which L$_{w}$ $\sim$ L$_{w,\star}$ are consistent with being wind-driven, bubbles where L$_{w}$ $>$ L$_{w,\star}$ require additional driving mechanisms other than only winds to account for their measured sizes the latter being also discussed in \citealt{relano05} for high-luminosity, extragalactic HII regions). In the case of the two growth-rate discrepancy bubbles, we suggest that the wind-driven expansion is slowed down due to surrounding higher-density material.

Here we have neglected the contribution of dust-processed radiation pressure, which L14 find to be comparable to the pressure of the warm ionised gas, and which could play a significant role in these two regions (particularly N44 C, which seems to be partially obscured by dusty features along its rim). With the MUSE data we cannot probe the dust-processed radiation pressure, and higher-resolution IR observations could help disentangle the various pressure terms at these spatial scales. This could for example be achieved with complementary JWTS/MIRI observations in the 5-28 $\mu$m range.

\section{Summary}\label{summary}

With data of the LMC regions N44 and N180 from the integral field spectrograph MUSE we simultaneously (identify and) classify the massive, feedback- driving O-type stars and derive the characteristics and kinematics of the feedback-driven bubbles in which the stars have formed. We analyse a total of 11 individual HII regions (with radii from 3 up to over 45 pc), which enables us to directly link the massive stellar population and feedback-related quantities of the regions. Our O star identification and classification algorithm correctly recovers all previously known O stars, and the MUSE data allows the identification of a total of 10 previously unidentified O-type stars.  

When comparing the observed luminosities with those expected from the stellar content of the individual regions, we find a deviation from the typically-used conversion between luminosity and ionising photon flux of a factor of 2. Whether this is due to the stellar model parameters used to infer the ionising photon flux from a given spectral type, or whether this is inherent to the low-metallicity environment of the LMC will be tested with future observations. 

We compute emission line ratio maps of the two HII region complexes and find a good agreement between derived electron density values and the literature. We investigate the ionisation structure with the O23 ratio and find that the highest degrees of ionisation are found in N44 D and N180 main. N44 main, the largest of the observed bubbles, hosting the oldest stellar population, is seemingly devoid of gas, which is reflected by a very low degree of ionisation. Furthermore, we compute oxygen abundances via the so-called strong line method and discuss the caveats of using this method for spatially resolved integral field data. We show the dependency of the derived values on the degree of ionisation, and conclude that integrated values agree (within errors) with the direct T$_{e}$-method. Radial velocities obtained from a Gaussian fitting routine agree well with the global dynamics of the LMC disk, as demonstrated when considering radial velocities as derived from Gaia on galactic scales.  

Feedback from the massive stellar population in the individual subregions is analysed by comparing the contribution of the direct radiation pressure, the pressure of the warm ionised gas and the pressure from stellar winds to the expansion of the HII regions. Overall, we find that the warm ionised gas and winds drive the expansion of the HII regions, with the ionised gas being the apparent dominant term. However, follow-up IR observations are needed to probe the contribution of dust-processed radiation pressure. Furthermore, we find that the computed photon escape fractions agree well with what is expected from literature. We also find that for the case of early-type O stars found in small ($<$ 10 pc) regions, where the high ionising flux produces high luminosities, radiation pressure is enhanced and higher star formation rates per unit are are measured. It would be important to compare this result for lower density bubbles with a comparable stellar content (i.e. photon flux budget) to derive the characteristic properties for which the various pressure terms dominate. 

We conclude that this work, together with a preliminary study of ionisation-induced photoevaporation of pillar-like structures in massive star-forming regions, demonstrates that integral field spectroscopy is a very powerful tool in the field of feedback from massive stars. However, disentangling the various feedback mechanisms only becomes possible with observations spanning a significant range of environments, as this allows to probe dependencies on metallicity, size, stellar and dust contents. We are currently planning a MUSE program targeting a statistically representative sample of HII regions in the LMC and SMC which, together with the preliminary data set presented here and in-hand MUSE data of HII regions in other nearby star-forming galaxies (e.g. McLeod et al., in prep.), will enable a robust analysis of feedback from massive stars.

\section*{Acknowledgements}
This research is supported by a Marsden Grant from the Royal Society of New Zealand (AFM). JMDK gratefully acknowledges funding from the German Research Foundation (DFG) in the form of an Emmy Noether Research Group (grant number KR4801/1-1) and from the European Research Council (ERC) under the European Union's Horizon 2020 research and innovation programme via the ERC Starting Grant MUSTANG (grant agreement number 714907). We thank M. Krumholz, P. Crowther and S. Glover for the useful discussions. This research made use of Astropy, a community-developed core Python package for Astronomy (Astropy Collaboration, 2018), as well as Astrodendro, a Python package to compute dendrograms of Astronomical data (http://www.dendrograms.org/), Pyspeckit \citep{pyspec}, and Spectral Cube (https://spectral-cube.readthedocs.io/).

\bibliographystyle{mn2e}
\bibliography{lmc_refs}



\appendix

\section{Uncertainty estimation}\label{errors}

In this section we briefly describe the source of uncertainties for the different quantities computed in this paper.

\begin{enumerate}
\item \textbf{Luminosity of the ionised gas L(H$\alpha$)}. The H$\alpha$ luminosity within a given radius is computed from the H$\alpha$ flux measured within the same radius, assuming a distance of 50 kpc to the LMC. The flux measurement is made from the extinction-corrected and continuum-subtracted H$\alpha$ emission line map. Given the good agreement between the MUSE and SHASSA H$\alpha$ fluxes (see main text), we compute L(H$\alpha$) by assuming a generous $\pm$10\% uncertainty on the measured flux. This is propagated to Q(H$\alpha$) and P$_{\mathrm{dir}}$. 
\item \textbf{Photon flux Q$_{0}$}. to convert the spectral type and luminosity class of the classified O-type stars we rely on the stellar parameters derived in \cite{martins05}. The main source of uncertainty for the photon flux is therefore the fact that these authors use solar metallicity. In lower-metallicity environment such as the LMC, stars are hotter and Lyman continuum fluxes are higher \citep{smith02}, but a similar set of calculations/calibrations are unavailable at present. The used photon fluxes are therefore most probably underestimated, and the derived $f_{\mathrm{esc}}$ a lower limit.
\item \textbf{Electron densities}. These two quantities are measured from the $n_{\mathrm{e}}$ and v$_{\mathrm{r}}$ maps from circular apertures centred on the regions of interest. The uncertainties stated in Table \ref{params} correspond to the standard deviation measured for each circular extraction. These are propagated to P$_{\mathrm{ion}}$, L$_{w}$ and P$_{w}$.

\end{enumerate}

\section{Stellar spectra}\label{n44spec}
\begin{figure*}
\centering
\mbox{
\subfloat[]{\includegraphics[scale=0.55]{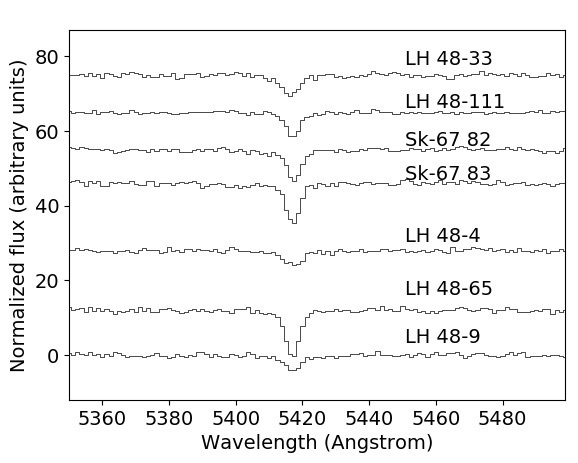}}
\subfloat[]{\includegraphics[scale=0.55]{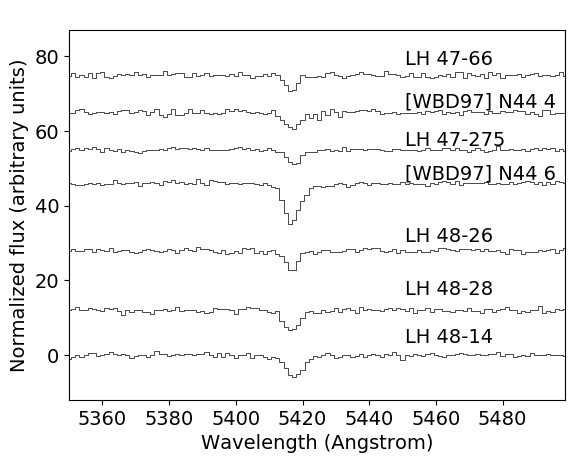}}}
\mbox{
\subfloat[]{\includegraphics[scale=0.55]{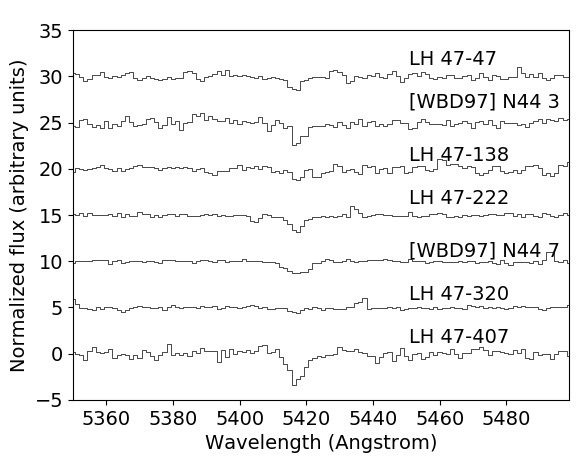}}
\subfloat[]{\includegraphics[scale=0.55]{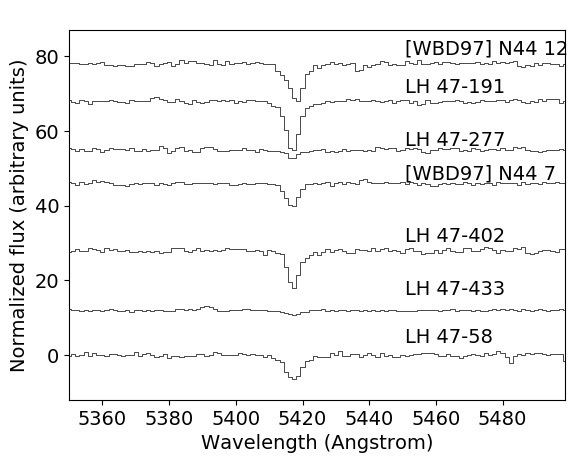}}}
\mbox{
\subfloat[]{\includegraphics[scale=0.55]{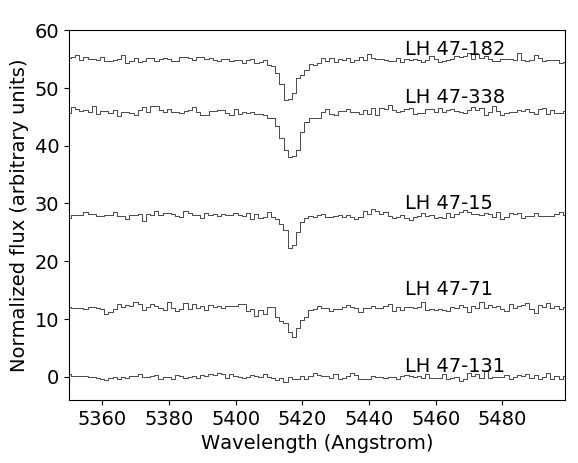}}
\subfloat[]{\includegraphics[scale=0.55]{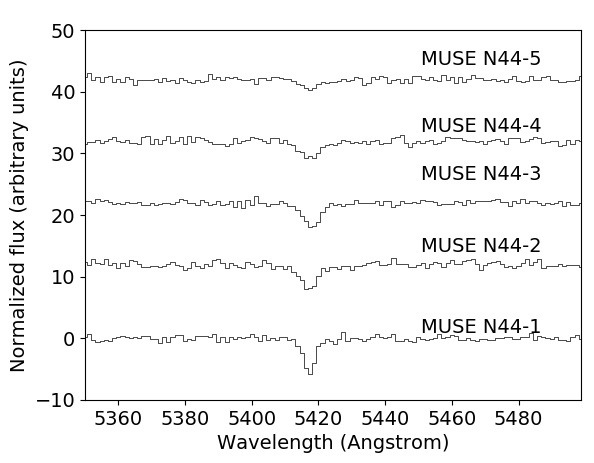}}}
\caption{He{\sc II} line of the O-type stars in N44. All detected lines have at least 2 data points above 3.5 times the noise.}
\label{n44heii_spec}
\end{figure*}

\begin{figure*}
\centering
\mbox{
\subfloat[]{\includegraphics[scale=0.55]{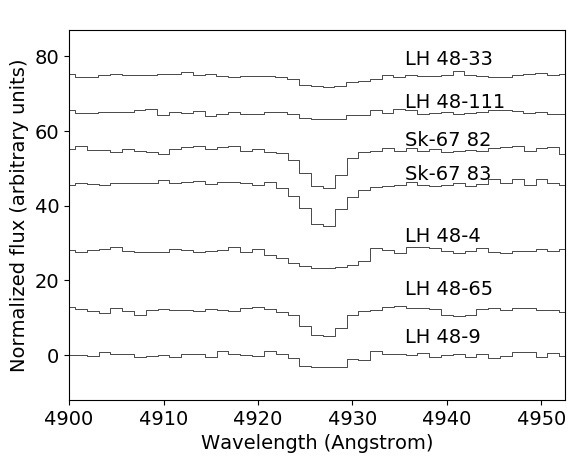}}
\subfloat[]{\includegraphics[scale=0.55]{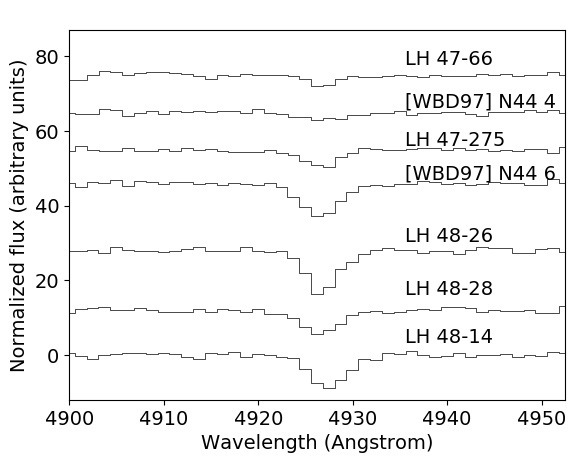}}}
\mbox{
\subfloat[]{\includegraphics[scale=0.55]{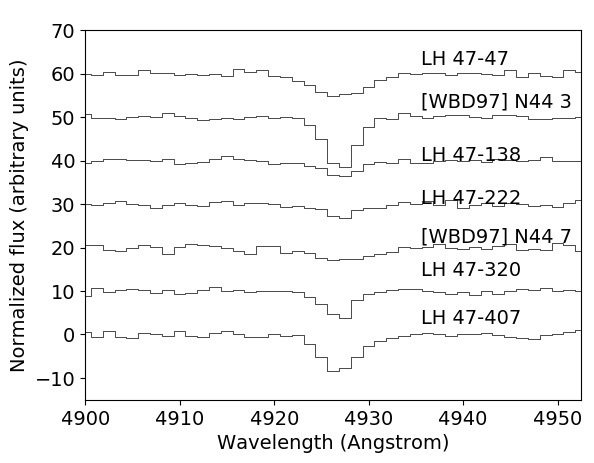}}
\subfloat[]{\includegraphics[scale=0.55]{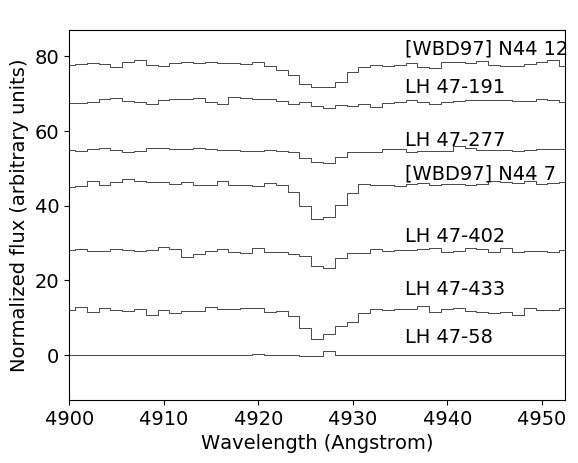}}}
\mbox{
\subfloat[]{\includegraphics[scale=0.55]{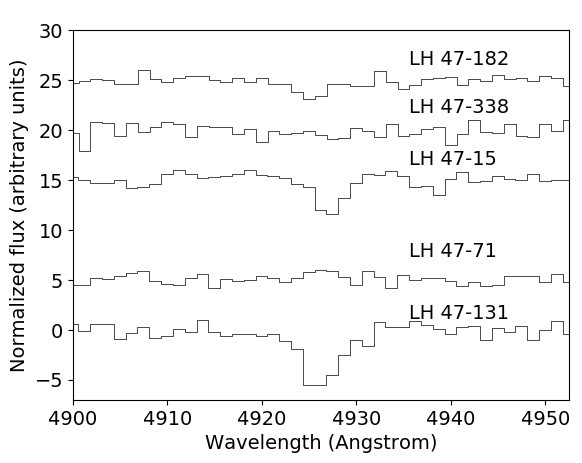}}
\subfloat[]{\includegraphics[scale=0.55]{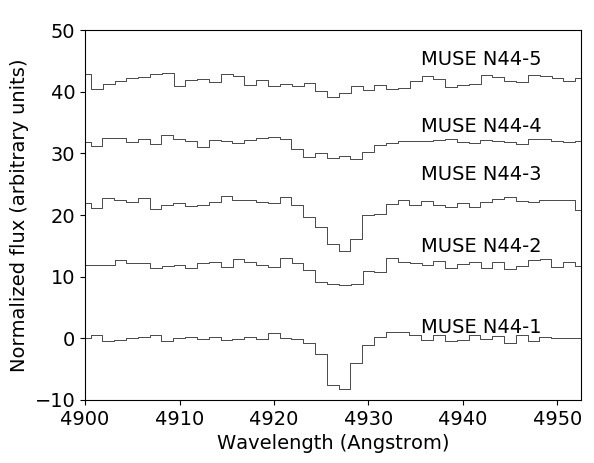}}}
\caption{He{\sc I} line of the O-type stars in N44.}
\label{n44hei_spec}
\end{figure*}

\begin{figure*}
\centering
\mbox{
\subfloat[]{\includegraphics[scale=0.55]{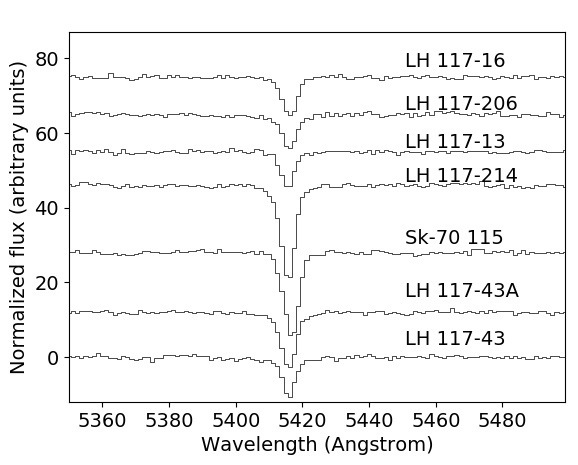}}
\subfloat[]{\includegraphics[scale=0.55]{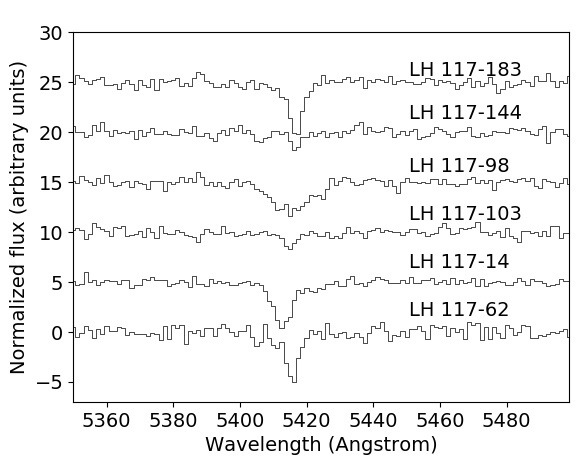}}}
\caption{He{\sc II} line of the 13 O-type stars in N180 main. All detected lines have at least 2 data points above 3.5 times the noise.}
\label{heii_spec}
\end{figure*}

\begin{figure*}
\centering
\mbox{
\subfloat[]{\includegraphics[scale=0.55]{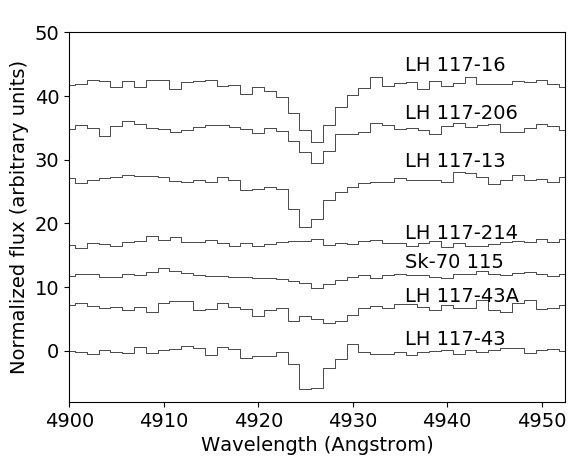}}
\subfloat[]{\includegraphics[scale=0.55]{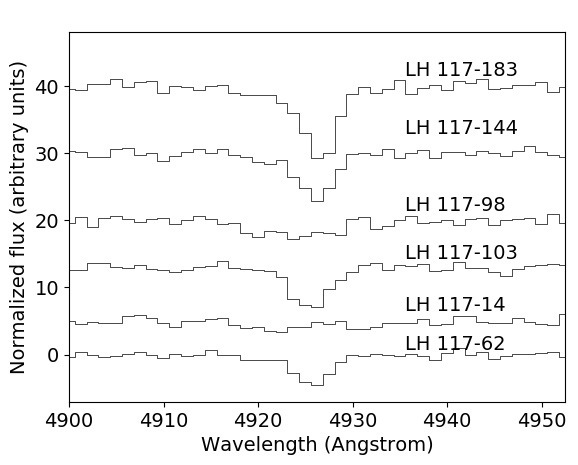}}}
\caption{He{\sc I} line of the N180 main O-type stars.}
\label{hei_spec}
\end{figure*}

\begin{figure*}
\centering
\mbox{
\subfloat[]{\includegraphics[scale=0.55]{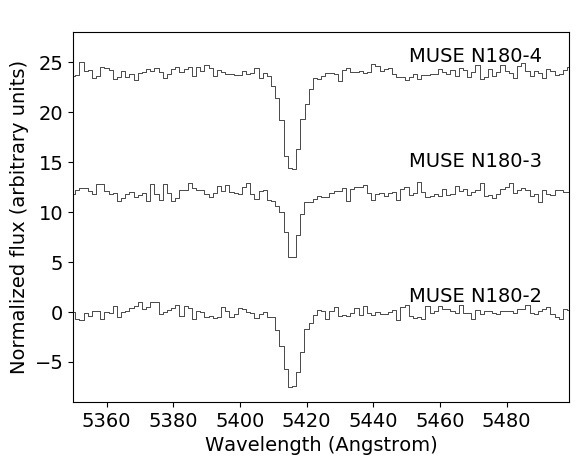}}
\subfloat[]{\includegraphics[scale=0.55]{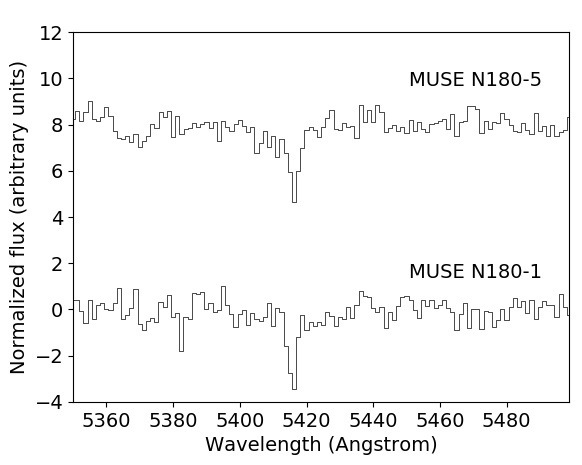}}}
\caption{He{\sc II} line of the newly identified O-type stars in N180. All detected lines have at least 2 data points above 3.5 times the noise.}
\label{heii_spec2}
\end{figure*}

Spectra of the identified O-type stars in N44 (Figures \ref{n44heii_spec} and \ref{n44hei_spec}) and N180 (Figures \ref{heii_spec}, \ref{hei_spec} and \ref{heii_spec2}). See main text Section \ref{stars_sec}.

\section{Wind luminosities}
We compute wind luminosities as expected from the spectral types of the stellar population within the single subregions according to model parameters given in \cite{muijres12}. For a given spectral type, the mass-loss rate and terminal velocity used to determine the wind luminosity are list in Table \ref{lw_star}, where terminal velocities are 2.6 times the escape velocity.

\begin{table}
\centering
\small
\begin{tabular}{lccc}
\hline
\hline
Spectral type & log($\dot M$) & $v_{\infty}$ & log(L$_{w,\star}$)\\
 & $M_{\odot}$/yr & km s$^{-1}$ & erg s$^{-1}$ \\
\hline
O3 V & -5.375 & 2740.4 & 37.00\\
O5 III & -5.491 & 2176.2 & 36.68 \\
O5 V & -5.829 & 2579.2 & 36.49 \\
O6.5 III & -5.902 & 2106.0 & 36.24 \\
O6.5 V & -6.427 & 2555.8 & 35.89 \\
O7 III & -6.016 & 2074.8 & 36.12 \\
O7 V & -6.624 & 2527.2 & 35.68 \\
O 7.5 I & -5.642 & 1859.0 & 36.39 \\
O7.5 V & -6.820 & 2493.4 & 35.47 \\
O8 III & -6.286 & 1996.8 & 35.81\\
O8 V & -7.019 &2454.4 & 35.26\\
O8.5 III & -6.409 & 1947.4 & 35.67\\
O8.5 V & -7.167 & 2399.8 & 35.09\\
O9 V &-7.374 & 2360.8 & 34.87\\
O9.5 I & -6.148 & 1593.8 & 35.57\\
O9.5 III & -6.646 &1843.3 & 35.38 \\
O9.5 V & -7.590 & 2319.2 & 34.64 \\
B0.5 III & -6.900 & 1500 & 34.95 \\
\hline
\hline
\end{tabular}
\\[1.5pt]
\caption{Mass-loss rates (column 2) and terminal velocities (column 3) from \citet{muijres12} used to derive the wind luminosity for a fixed spectral type (column 4).}
\label{lw_star}
\end{table}

\section{Radial intensity profiles}
\begin{figure*}
\centering
\mbox{
\subfloat[]{\includegraphics[scale=0.55]{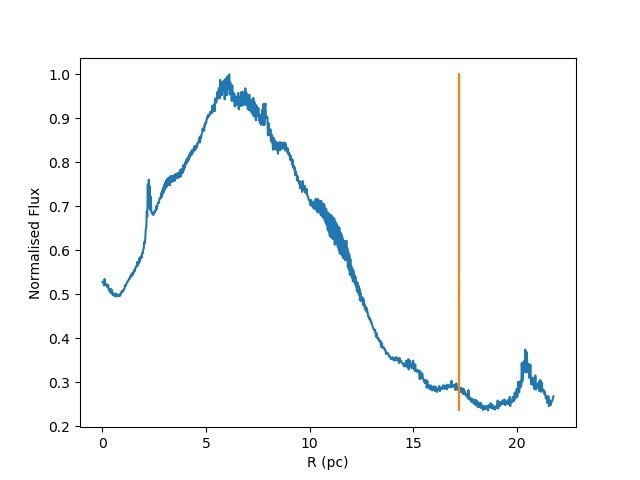}}
\subfloat[]{\includegraphics[scale=0.55]{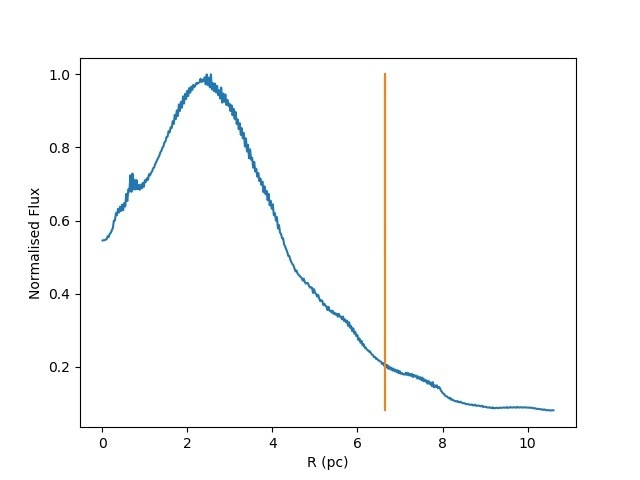}}}
\mbox{
\subfloat[]{\includegraphics[scale=0.55]{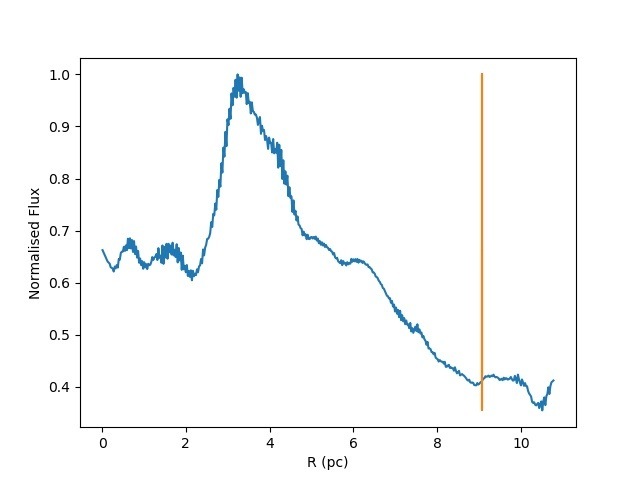}}
\subfloat[]{\includegraphics[scale=0.55]{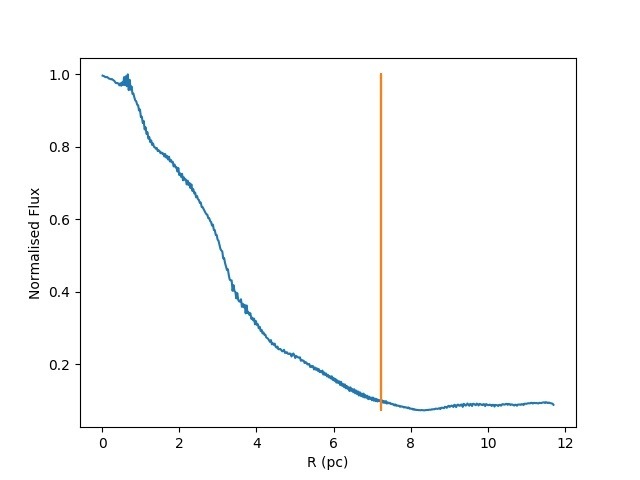}}}
\mbox{
\subfloat[]{\includegraphics[scale=0.55]{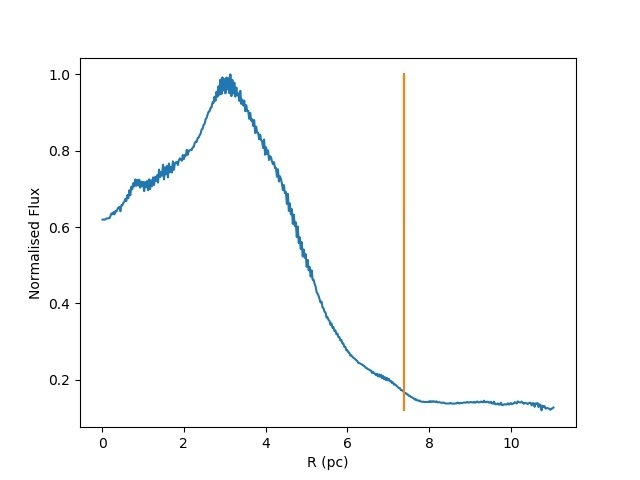}}
\subfloat[]{\includegraphics[scale=0.55]{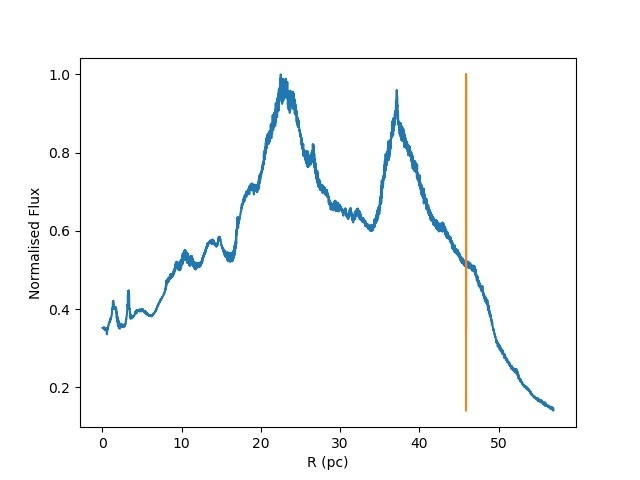}}}
\caption{Radial intensity profiles for N44 A (a), N44 B (b), N44 B1 (c) N44C (d), N44 D (e) and N44 main (f). Vertical lines indicate the radius which encompasses 90\% of the measured flux.}
\label{n44_profiles}
\end{figure*}

\begin{figure*}
\centering
\mbox{
\subfloat[]{\includegraphics[scale=0.55]{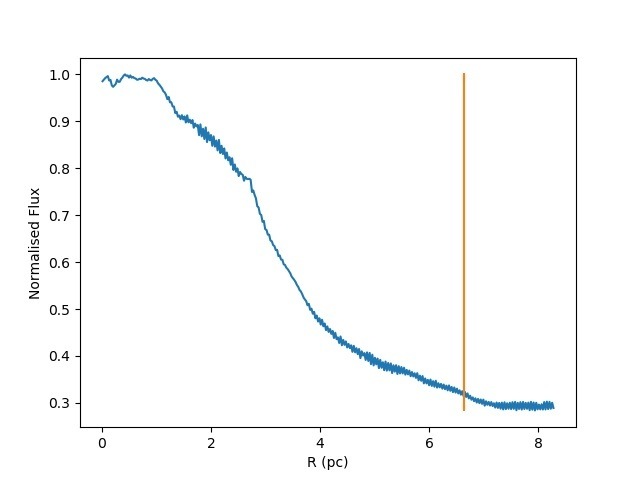}}
\subfloat[]{\includegraphics[scale=0.55]{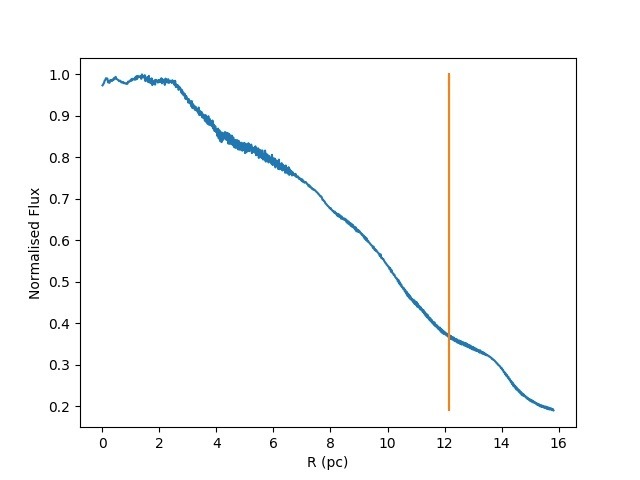}}}
\mbox{
\subfloat[]{\includegraphics[scale=0.55]{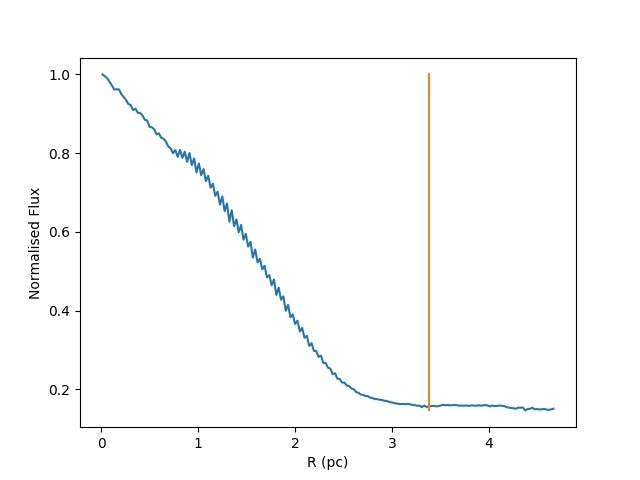}}
\subfloat[]{\includegraphics[scale=0.55]{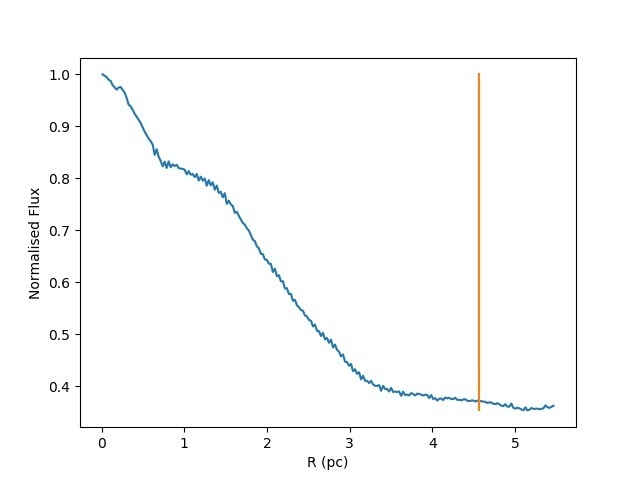}}}
\mbox{
\subfloat[]{\includegraphics[scale=0.55]{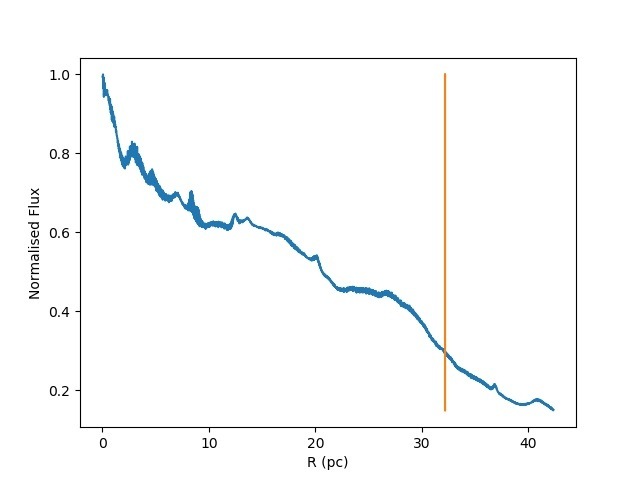}}}
\caption{Radial intensity profiles for N180 A (a), N180 B (b), N180C (c) and N180 main (e). Vertical lines indicate the radius which encompasses 90\% of the measured flux.}
\label{n180_profiles}
\end{figure*}

Sizes are determined as the radii encompassing 90\% of the measured H$\alpha$ flux. Radial profiles are shown in Figures \ref{n44_profiles} and \ref{n180_profiles}.

\bsp

\label{lastpage}

\end{document}